\documentclass[prb,twocolumn]{revtex4-2}
\usepackage{graphicx} 

\usepackage{amsmath,amsthm,amssymb,mathrsfs}
\usepackage{bm}
\usepackage{color}

\begin{document}

\title{Computational capability for physical reservoir computing using a spin-torque oscillator with two free layers}

\author{Terufumi Yamaguchi$^{1}$}
\email{terufumi.yamaguchi@riken.jp}
\author{Sumito Tsunegi$^{2,3}$}
\author{Kohei Nakajima${}^{4}$}
\author{Tomohiro Taniguchi$^{2}$}
\email{tomohiro-taniguchi@aist.go.jp}

\affiliation{ 
${}^{1}$RIKEN, Wako, Saitama 351-0198, Japan, 
\\
${}^{2}$National Institute of Advanced Industrial Science and Technology (AIST), Research Center for Emerging Computing Technologies, Tsukuba 305-8568, Japan,
\\
${}^{3}$PRESTO, Japan Science and Technology Agency (JST), Saitama 332-0012, Japan, 
\\
 ${}^{4}$Graduate School of Information Science and Technology, The University of Tokyo, Bunkyo-ku, 113-8656 Tokyo, Japan. 
}

\date{\today}%

\begin{abstract}
A numerical analysis on the computational capability of physical reservoir computing utilizing a spin-torque oscillator with two free layers is reported. 
Conventional spintronics devices usually consist of two ferromagnets, where the direction of magnetization in one layer, called the free layer, can move while that of the other, the reference layer, is fixed. 
Recently, however, devices with two free layers, where the reference layer is replaced by another free layer, have been developed for various practical applications. 
Adding another free layer drastically changes the dynamical response of the device through the couplings via the spin-transfer effect and the dipole magnetic field. 
A numerical simulation of the Landau-Lifshitz-Gilbert equation and a statistical analyses of the Lyapunov exponent and the synchronization index reveal the appearance of an amplitude-modulated oscillation and chaos in the oscillators with two free layers. 
Such complex dynamics qualitatively change the computational capability of physical reservoir computing because the computational resource is dynamics of the physical system. 
An evaluation of the short-term memory capacity clarifies that oscillators with two free layers have a larger capacity than those of conventional oscillators. 
An enhancement in capacity near the edge of echo state property, i.e., the boundary between zero and finite synchronization index, is also found.   
\end{abstract}

\maketitle


\section{Introduction}
\label{sec:Introduction}

Recent developments in spintronics applications, such as brain-inspired computing \cite{grollier20}, have led to a variety of device structures and materials \cite{locatelli14}. 
For example, giant-magnetoresistive (GMR) \cite{baibich88,binasch89,pratt91} and tunnel-magnetoresistive (TMR) \cite{juliere75,maekawa82,miyazaki95,moodera95,yuasa04JJAP,parkin04,yuasa04} structures that include ferromagnets, called free and reference layers, have been used in magnetic sensors and memories \cite{dieny16}. 
The magnetization in the free layer can change its direction when a magnetic field and/or electric current is applied to it \cite{slonczewski96,berger96}.
On the other hand, the reference layer often consists of two ferromagnets separated by a thin nonmagnetic spacer, and the antiferromagnetic interlayer exchange coupling between them strongly fixes their magnetization directions.  
Moreover, GMR and TMR devices with two free layers, where the reference layer is replaced by another ferromagnet without pinning effects, have recently been investigated for new applications such as high-density magnetic recording \cite{zhou19}, probabilistic computing \cite{camsari21}, and millimeter-wave generator \cite{kurokawa22}. 
In such devices, the coupled dynamics of the magnetizations of the two free layers that arise through the via spin-transfer effect and magnetic dipole field provide new functionalities. 


A critical difference in the magnetization dynamics between a GMR/TMR device with a single free layer and a device with two free layers is the appearance of chaos in the latter structure because of the increased dynamical degrees of freedom \cite{kudo06,taniguchi19,taniguchi20,matsumoto19}. 
In particular, devices with two free layers might be applicable to physical reservoir computing \cite{maas02,jaeger04,verstraeten07,nakajima20,nakajima21}, which is another new application of spintronics technology. 
Physical reservoir computing is a kind of recurrent neural network in which a reservoir, which is a physical nonlinear system, performs a computational task; for example, a spin-torque oscillator (STO) has been applied to the task of human voice recognition \cite{torrejon17}. 
Since physical reservoir computing utilizes dynamical output signals from a physical system as a computational resource, the recent research has viewed the relation between the computational capability and the dynamical state of the physical system to be of central importance \cite{nakajima21}. 
Such investigations in spintronics \cite{akashi20,akashi22} have recently focused on computing near the chaotic state because the edge of chaos sometimes provides a boundary of high computational capability \cite{bertschinger04,nakayama16}. 
Therefore, due to the appearance of chaos, the computational performance of physical reservoir computing by using an STO with two free layers might be also different from that of an STO with single free layer. 


\begin{figure*}
\centerline{\includegraphics[width=2.0\columnwidth]{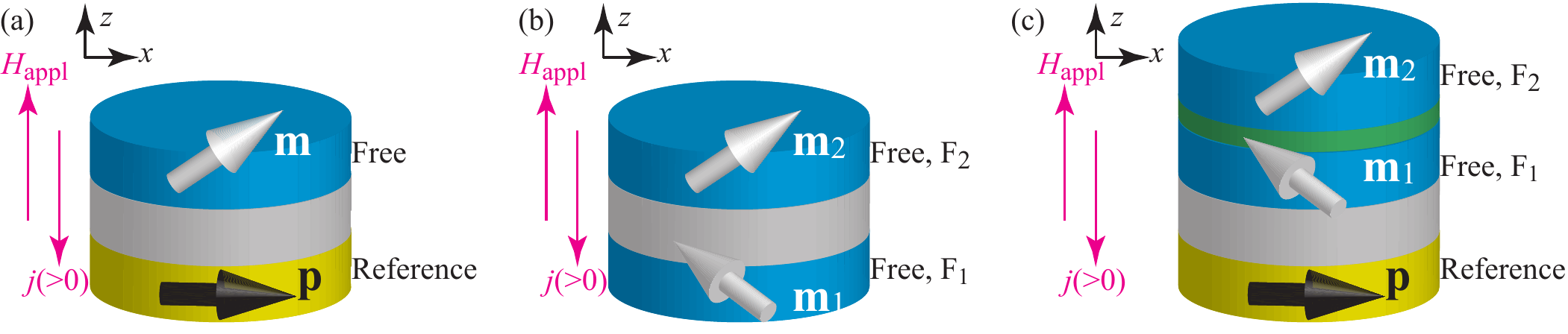}}
\caption{
         Schematic illustrations of the device structures studied in this paper. 
         The unit vectors pointing in the magnetization directions of the free and reference layers are denoted as $\mathbf{m}$ and $\mathbf{p}$. 
         A suffix $i=1,2$ is added to $\mathbf{m}$ when the device includes two free layers. 
         (a) Conventional GMR/TMR structure consisting of one free and one reference layer. 
              The output signal through the GMR/TMR effect is proportional to $\mathbf{m}\cdot\mathbf{p}=m_{x}$.  
         (b) Device consisting of two free layers, F${}_{1}$ and F${}_{2}$. 
              The output signal is proportional to $\mathbf{m}_{1}\cdot\mathbf{m}_{2}$. 
         (c) Device consisting of two free layers, F${}_{1}$ and F${}_{2}$, and one reference layer. 
              The output signal is proportional to $\mathbf{m}_{1}\cdot\mathbf{p}$, assuming that the GMR/TMR effect between F${}_{1}$ and reference layer is dominant. 
         The directions of positive current density and the external magnetic field are indicated by arrows. 
         \vspace{-3ex}}
\label{fig:fig1}
\end{figure*}


In this paper, we evaluate the computational performance of STOs with two free layers by performing numerical simulations of the Landau-Lifshitz-Gilbert (LLG) equation. 
We consider the three structures, schematically shown in Fig. \ref{fig:fig1}. 
The first one, in Fig. \ref{fig:fig1}(a), is a conventional GMR or TMR system, which consists of one free layer and one reference layer. 
The second structure, in Fig. \ref{fig:fig1}(b), consists of two free layers, where the magnetizations of both layers can change their directions through the spin-transfer effect and the dipole coupling. 
The third structure, in Fig. \ref{fig:fig1}(c), includes three ferromagnets; two are free layers and one is a reference layer. 
We find that the first structure shows a saturation to a fixed point, while the second and third structures show a wide variety of dynamics, such as amplitude-modulated oscillation and chaos. 
These dynamical states are classified systematically by measuring the Lyapunov exponent and synchronization index, which are measures for identifying chaotic dynamics and the echo state property. 
In addition, the short-term memory capacity is evaluated as a figure of merit of the computational capability. 
It is found that the STOs with two free layers have larger capacities than that of the STO with a single free layer. 
An enhancement in capacity near the edge of the echo state property, i.e., the boundary between zero and finite synchronization index, is also observed in the STOs with two free layers. 


The paper is organized as follow. 
Section \ref{sec:STO with single free layer} examines the dynamical state of the magnetization, short-term memory capacity, Lyapunov exponent, and synchronization index of an STO with a single free layer is studied, while Secs. \ref{sec:STO consisting of two free layers} and \ref{sec:STO consisting of two free and one reference layers} examine those features of an STO consisting of two free layers and an STO with two free layers and one reference layer. 
Section \ref{sec:Conclusions} is the conclusion. 



\section{STO with single free layer}
\label{sec:STO with single free layer} 

Here, we analyze the dynamics of a conventional STO consisting of a free and reference layer by the LLG equation and summarize the methods of evaluating the short-term memory capacity, Lyapunov exponent, and synchronization index. 
We use the macrospin LLG equation based on the model in Ref. \cite{taniguchi19}, where the accuracy of the macrospin model was verified by the comparison with the experiment \cite{zhou19}. 
The results will be compared to those of STOs with two free layers in Secs. \ref{sec:STO consisting of two free layers} and \ref{sec:STO consisting of two free and one reference layers}. 


\subsection{LLG equation of STO with single free layer}
\label{sec:LLG equation of STO with single free layer}

The STO is schematically shown in Fig. \ref{fig:fig1}(a). 
The unit vectors pointing in the magnetization direction of these layers are denoted as $\mathbf{m}$ and $\mathbf{p}$, respectively. 
The magnetization dynamics in the free layer are described by the LLG equation, 
\begin{equation}
  \frac{d \mathbf{m}}{dt}
  =
  -\gamma
  \mathbf{m}
  \times
  \mathbf{H}
  +
  \gamma
  H_{\rm s}
  \mathbf{m}
  \times
  \left(
    \mathbf{p}
    \times
    \mathbf{m}
  \right)
  +
  \alpha
  \mathbf{m}
  \times
  \frac{d \mathbf{m}}{d t}
  \label{eq:LLG_single_free}
\end{equation}
where the magnetic field $\mathbf{H}$ consists of the shape magnetic anisotropy field and an external magnetic field $H_{\rm appl}$ applied along the perpendicular ($z$) direction, 
\begin{equation}
  \mathbf{H}
  =
  \begin{pmatrix}
    -4\pi M N_{x} m_{x} \\
    -4\pi M N_{y} m_{y} \\
    H_{\rm appl} - 4\pi M N_{z} m_{z}
  \end{pmatrix}.
  \label{eq:field_single_free}
\end{equation}
The demagnetization coefficients are denoted as $N_{\ell}$ ($N_{x}+N_{y}+N_{z}=1$). 
The spin-transfer torque strength is 
\begin{equation}
  H_{\rm s}
  =
  \frac{\hbar \eta j}{2e(1 + \lambda \mathbf{m}\cdot\mathbf{p}) Md}, 
  \label{eq:H_s}
\end{equation}
where $M$ and $d$ are the saturation magnetization and the thickness of the free layer. 
The spin polarization of the current density $j$ is $\eta$, while $\lambda$ provides the spin-transfer torque asymmetry \cite{slonczewski96}. 
A positive current corresponds to a flow of electrons from the reference to the free layer. 
The values of the parameters are $M=1300$ emu/cm${}^{3}$, $\eta=0.30$, $\lambda=\eta^{2}$, $d=2$ nm, $\gamma=1.764\times 10^{7}$ rad/(Oe s), $\alpha=0.010$, and $H_{\rm appl}=1.0$ kOe.  
The demagnetization coefficients are \cite{tandon03,taniguchi18JMMM} 
\begin{widetext}
\begin{equation}
  N_{z}
  =
  \frac{1}{\tau}
  \left\{
    \frac{3}{4\pi}
    -
    \frac{3}{4\pi}
    \sqrt{1+\tau^{2}}
    \left[
      \tau^{2}
      \mathsf{K}
      \left(
        \frac{1}{\sqrt{1+\tau^{2}}}
      \right)
      +
      \left(
        1
        -
        \tau^{2}
      \right)
      \mathsf{E}
      \left(
        \frac{1}{\sqrt{1+\tau^{2}}}
      \right)
    \right]
    +
    \tau
  \right\},
\end{equation}
\end{widetext}
and $N_{x}=N_{y}=(1-N_{z})/2$, where $\tau=d/(2r)$ and $r=50$ nm is the radius of the free layer. 
Here, we assume that the layer has a cylinder shape. 
The first and second kinds of complete elliptic integral with the modulus $k$ are $\mathsf{K}(k)=\int_{0}^{\pi/2}d\phi/\sqrt{1-k^{2}\sin^{2}\phi}$ and $\mathsf{E}(k)=\int_{0}^{\pi/2}d \phi \sqrt{1-k^{2}\sin^{2}\phi}$. 
Furthermore, we assume that the magnetization in the reference layer points to an in-plane ($x$) direction, i.e., $\mathbf{p}=\hat{\mathbf{e}}_{x}$, where $\hat{\mathbf{e}}_{\ell}$ is the unit vector in the $\ell$ ($\ell=x,y,z$) direction.


\begin{figure}
\centerline{\includegraphics[width=1.0\columnwidth]{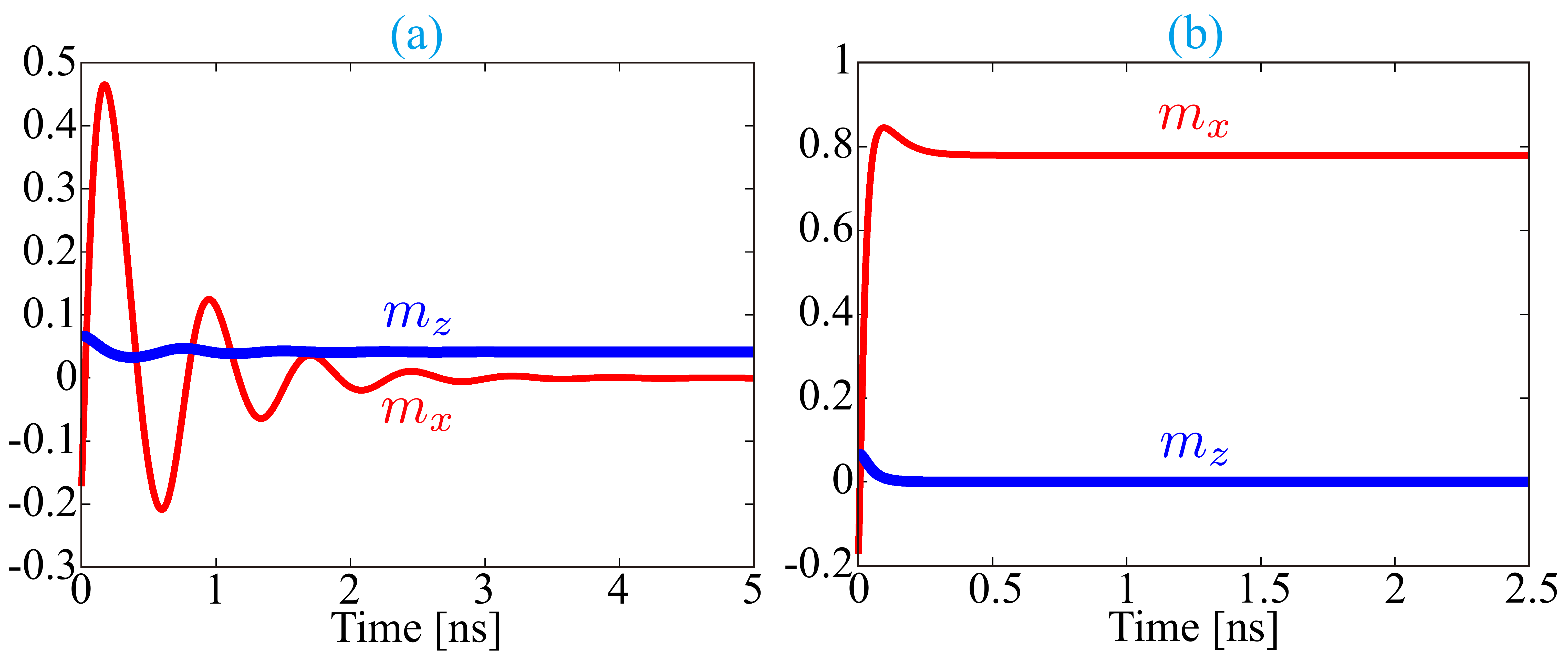}}
\caption{
            Examples of dynamics of the in-plane ($m_{x}$) and out-of-plane ($m_{z}$) components of magnetization for current densities of (a) $100$ and (b) $450$ MA/cm${}^{2}$ in an STO with a single free layer. 
         \vspace{-3ex}}
\label{fig:fig2}
\end{figure}


Figures \ref{fig:fig2}(a) and \ref{fig:fig2}(b) show typical magnetization dynamics for low ($j_{0}=100$ MA/cm${}^{2}$) and high ($j_{0}=450$ MA/cm${}^{2}$) current densities. 
Starting from the steady state in the absence of a current, where $m_{z}=H_{\rm appl}/(4\pi M N_{z})$, the magnetization eventually saturates to fixed points. 
When the magnitude of the current is small, the magnetization points to the direction orthogonal to the $x$ axis, i.e., $m_{x}$ becomes zero, for the following reason. 
The steady point is determined by the condition $d \mathbf{m}/dt=\bm{0}$, which, in the present case, means that $\mathbf{H}-H_{\rm s}\mathbf{p}\times \mathbf{m} =\bm{0}$. 
Since $N_{z}\simeq 1$, $N_{x},N_{y} \ll 1$ and $4\pi MN_{z} \gg H_{\rm appl}$, the magnetic field $\mathbf{H}$ mainly points in the $z$ direction. 
Then, because $\mathbf{p}$ points to the $x$ direction, the magnetization $\mathbf{m}$ should point in the $y$ direction in order to satisfy $\mathbf{H}-H_{\rm s}\mathbf{p}\times \mathbf{m} =\bm{0}$. 
Thus, the magnetization saturates to a fixed point of $|m_{y}|\simeq 1$, as shown in Fig. \ref{fig:fig2}(a). 
When the current magnitude becomes large, the magnetization moves to different fixed points, where, because of the strong spin-transfer torque, the magnetization becomes close to parallel or antiparallel to the magnetization in the reference layer, depending on the sign of the current. 
In the present system, a positive current prefers the parallel alignment of $\mathbf{m}$ and $\mathbf{p}$, and therefore, the magnetization saturates to a fixed point with $m_{x}>0$ for the case shown in Fig. \ref{fig:fig2}(b). 


In the next section, we study magnetization dynamics in the presence of a series of random input signals. 
The dynamical response to such input signals was used in a recognition task in physical reservoir computing \cite{fujii17,tsunegi18}. 
There, it is necessary to specify the quantity to be used as the output signal. 
The output signal from the present STO depends on the magnetization direction in the free layer along the direction of the magnetization in the reference layer, i.e., $\mathbf{m}\cdot\mathbf{p}$, due to the GMR/TMR effect. 
Therefore, we will choose to use $\mathbf{m}\cdot\mathbf{p}=m_{x}$ as the output signal used for computing.



\begin{figure}
\centerline{\includegraphics[width=1.0\columnwidth]{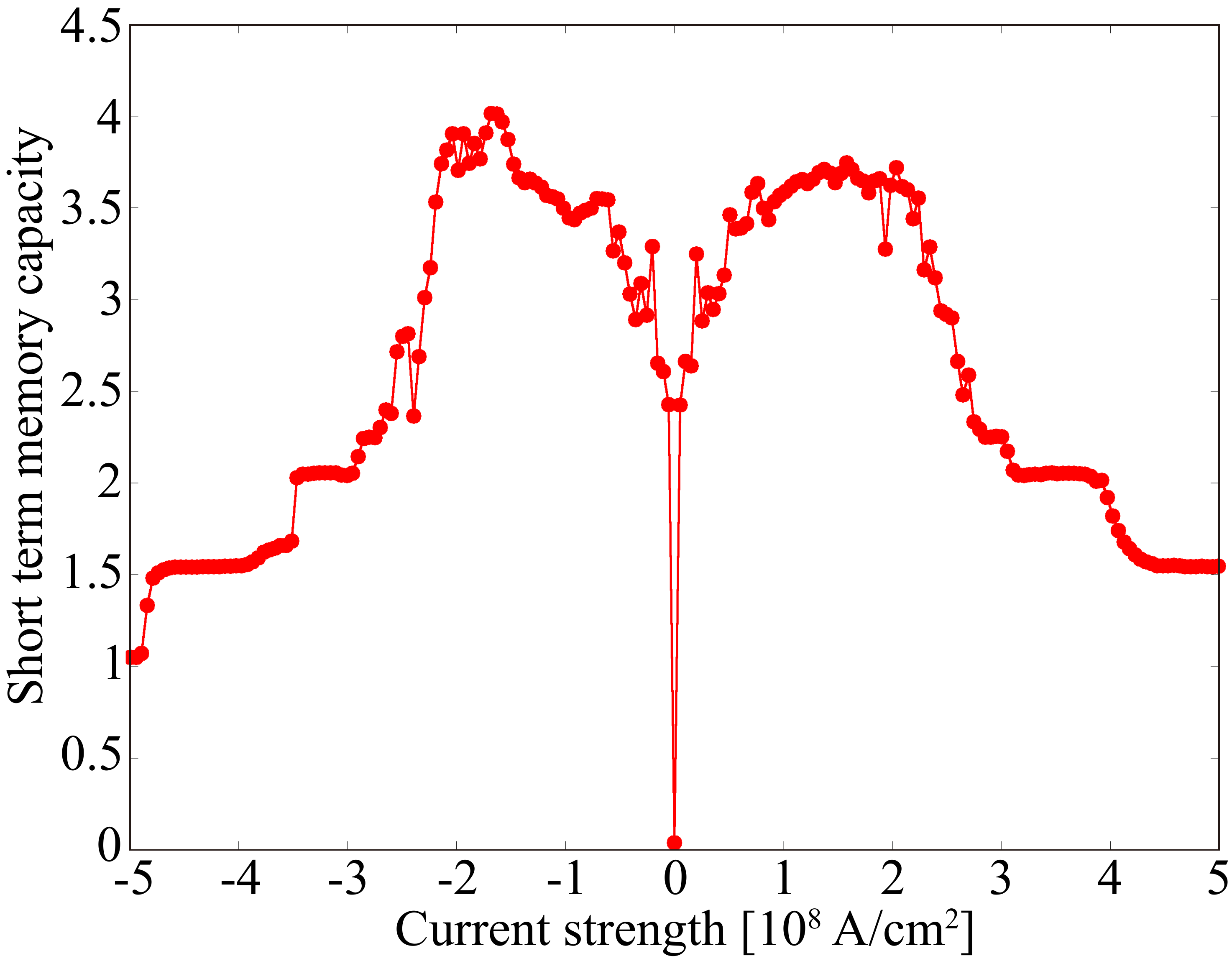}}
\caption{
            Dependence of short-term memory capacity on current density for an STO with a single free layer. 
         \vspace{-3ex}}
\label{fig:fig3}
\end{figure}


\subsection{Short-term memory capacity}
\label{sec:Short-term memory capacity}

We will quantify the computational capability of the STO by its short-term memory capacity. 
The short-term memory capacity corresponds to, roughly speaking, the number of input data a physical reservoir can recognize. 
Therefore, a large short-term memory capacity corresponds to a high computational performance; see also Sec. \ref{sec:Conclusions}, where the relation between the short-term memory capacity and the total computational capability is briefly explained. 
In the present paper, we suppose a binary pulse-input signal $b_{k}=0,1$ ($k=1,2,\cdots$) \cite{fujii17,tsunegi18,yamaguchi20} with a pulse width of $t_{\rm p}$, which is added to the current density as 
\begin{equation}
  j
  =
  j_{0}
  \left(
    1
    +
    \nu 
    b_{k}
  \right),
  \label{eq:input_current_def}
\end{equation}
where the dimensionless parameter $\nu$ quantifies the strength of the input signal, while $j_{0}$ is the current density for $b_{k}=0$. 
In the following, we refer $j_{0}$ as the current density for simplicity, while $j$ is referred as the total current density. 
The suffix $k$ distinguishes the order of the input data. 
The values of $\nu$ and $t_{\rm p}$ are assumed to be $0.2$ and $1.0$ ns. 
The input signal to the current changes the magnetization dynamics through modulation of the spin-transfer torque. 
The details of the evaluation method are summarized in Appendix \ref{sec:AppendixA} (see also, for example, Ref. \cite{tsunegi18}). 

Figure \ref{fig:fig3} summarizes the current dependence of the short-term memory capacity. 
High performance is obtained when the magnitude of the current is relatively small and the magnetization points in a direction orthogonal to the $x$ axis; see also Fig. \ref{fig:fig2}. 
The maximum value of the short-term memory capacity is $4.02$ at $j_{0}=-168$ MA/cm${}^{2}$. 
The step-like behavior in the large current region is similar to that observed in a different STO \cite{yamaguchi20srep}. 
When the current density $j_{0}$ is zero, the short-term memory capacity is zero because, according to Eq. (\ref{eq:input_current_def}), the total current is zero even if $b_{k}$ is finite, and thus, the input signal does not cause any change in the magnetization state. 
In the following, we evaluate the Lyapunov exponent of the STO and show that such a current dependence of the short-term memory capacity relates to a relaxation time of the magnetization to the fixed point. 


\subsection{Edges of chaos and echo state property}
\label{sec:Edges of chaos and echo state property}

Since we aim to reveal the relation between the computational capability and dynamical state and quantify this capability by the short-term memory capacity, it becomes necessary to introduce quantities distinguishing the dynamical state. 
We will use the Lyapunov exponent \cite{yamaguchi19,taniguchi19PRB,taniguchi20} and the synchronization index \cite{akashi20} for this purpose. 
While their evaluation methods are described in Secs. \ref{sec:Lyapunov exponent} and \ref{sec:Synchronization index}, here, let us briefly explain their roles and differences. 


The Lyapunov exponent is the inverse of the time scale of the expansion of the distance between two solutions to the LLG equation with slightly different initial conditions \cite{strogatz01}. 
While the number of the Lyapunov exponents is the same with that of the dynamical degree of freedom, the Lyapunov exponent focused in this paper corresponds to the largest Lyapunov exponent; see also Appendix \ref{sec:AppendixB}. 
When the Lyapunov exponent is negative (positive), the difference between the states decreases (increases) relative to that of the initial state as time increases. 
Therefore, a system with a positive Lyapunov exponent has a high sensitivity to its initial state. 
Typical dynamics for negative, zero, and positive Lyapunov exponents are a saturation to a fixed point, a limit-cycle oscillation, and chaos, respectively \cite{strogatz01}. 


The Lyapunov exponent often changes value and even sign when the system parameters are changed \cite{strogatz01}. 
The boundary between a zero and positive Lyapunov exponents is called the edge of chaos. 
There are methods to evaluate the Lyapunov exponent from numerical simulations of the equation of motion \cite{shimada79,muller95,alligood97,ott02,kanno14}.  
There are also statistical methods that evaluate the Lyapunov exponent from time-series data \cite{farmer82,wolf85,rosenstein93,kantz94}. 
While statistical analyses have frequently been used for analyzing experimental data, they have restrictions; for example, some \cite{rosenstein93,kantz94} can only evaluate positive Lyapunov exponents, i.e., only the edge of chaos can be identified, while the boundary between negative and zero Lyapunov exponent cannot be estimated. 
The present paper quantifies the Lyapunov exponent through numerical simulations; thus, it identifies all possible values of the exponents, i.e., negative, zero, and positive.


The synchronization index \cite{akashi20} is a long-time average of the distance between two independent samples of the magnetization which obey the same LLG equation but have slightly different initial conditions. 
The synchronization index becomes zero when the dynamical state becomes independent of the initial state as time goes on. 
In an autonomous system, the synchronization index tends to be zero when the dynamics saturate to a fixed point. 
An example is magnetization switching, where the magnetization eventually points in a certain direction.  
On the other hand, when the magnetization is, for example, in an auto-oscillation state, the difference between the states in the oscillation phases will never decrease below the initial difference because the two magnetizations oscillate with the same frequency. 
In such a case, the synchronization index remains finite. 


The situation changes when a time-dependent signal is injected. 
For example, in the case of a periodic input signal, e.g., as in forced synchronization \cite{quinsat11}, the phase of the magnetization is fixed with respect to that of the periodic input signal. 
In this case, the synchronization index becomes zero even if the magnetizations are in an oscillating state. 
Another example of a time-dependent input signal is a random signal, which is used in physical reservoir computing, as mentioned in Sec. \ref{sec:Short-term memory capacity}. 
In this case, nonlinear oscillators often show noise-induced synchronization \cite{mainen95,toral01,teramae04,goldobin05,nakao07,imai22}, and the oscillating state of the magnetization eventually becomes independent of the initial state. 
Then, the synchronization index becomes zero. 
In particular, noise-induced synchronization is of interest in physical reservoir computing because this synchronization behavior in at physical system guarantees the echo state property \cite{yildiz12}.  
The echo state property is a necessary condition guaranteeing the computational reproducibility, wherein the dynamical state of the physical reservoir becomes independent of the initial state by injecting random input signals as washout (see also Appendix \ref{sec:AppendixA}); therefore, the physical reservoir always provides the same answer for the same task. 
The boundary between zero and finite synchronization indexes will be called the edge of the echo state property. 

One might imagine that the edge of the echo state property can be identified from the boundary between negative and zero Lyapunov exponent. 
Here, saturation to a fixed point is an example of magnetization dynamics corresponding to a negative Lyapunov exponent and auto-oscillation to a zero Lyapunov exponents; according to the above discussion, saturation to a fixed point should lead to a zero synchronization index and auto-oscillation to a nonzero index. 
Therefore, one might imagine that it is unnecessary to evaluate the synchronization index. 
However, in other cases, knowing the Lyapunov exponent is not sufficient to clarify the edge of the echo state property for the following reasons. 
First, the Lyapunov exponent in this study is, strictly speaking, the maximum Lyapunov exponent, which corresponds to the expansion rate in a direction along which the difference between the initial states grows the most.  
Second, in many cases, only some of the dynamical variables are used for computing. 
For example, the dynamical variable used for computing sometimes has the echo state property even though the maximum Lyapunov exponent is zero. 
Such an example will be shown in Sec. \ref{sec:STO consisting of two free layers} below.

From the above it is clear that the Lyapunov exponent and the synchronization index are similar but slightly different quantities. 
The former determines the edge of chaos, while the latter determines the edge of the echo state property. 
A periodic oscillation state is an example of a dynamical state separating these edges, which does not have the echo state property and is non-chaotic. 
While the computational capability of the optical physical reservoir computing presented in Ref. \cite{nakayama16} is maximized at the edge of chaos, Ref. \cite{nakajima21} argues that chaos is not necessary for the computational capability to be enhanced; rather, the edge of the echo state property often corresponds to an optimization condition.
Moreover, although these two edges might overlap in some cases \cite{akashi20}, this is not guaranteed to happen in all cases. 
In the present paper, therefore, we estimated these edges from the Lyapunov exponent and the synchronization index, and studied their relation to the computational capability. 



\begin{figure}
\centerline{\includegraphics[width=1.0\columnwidth]{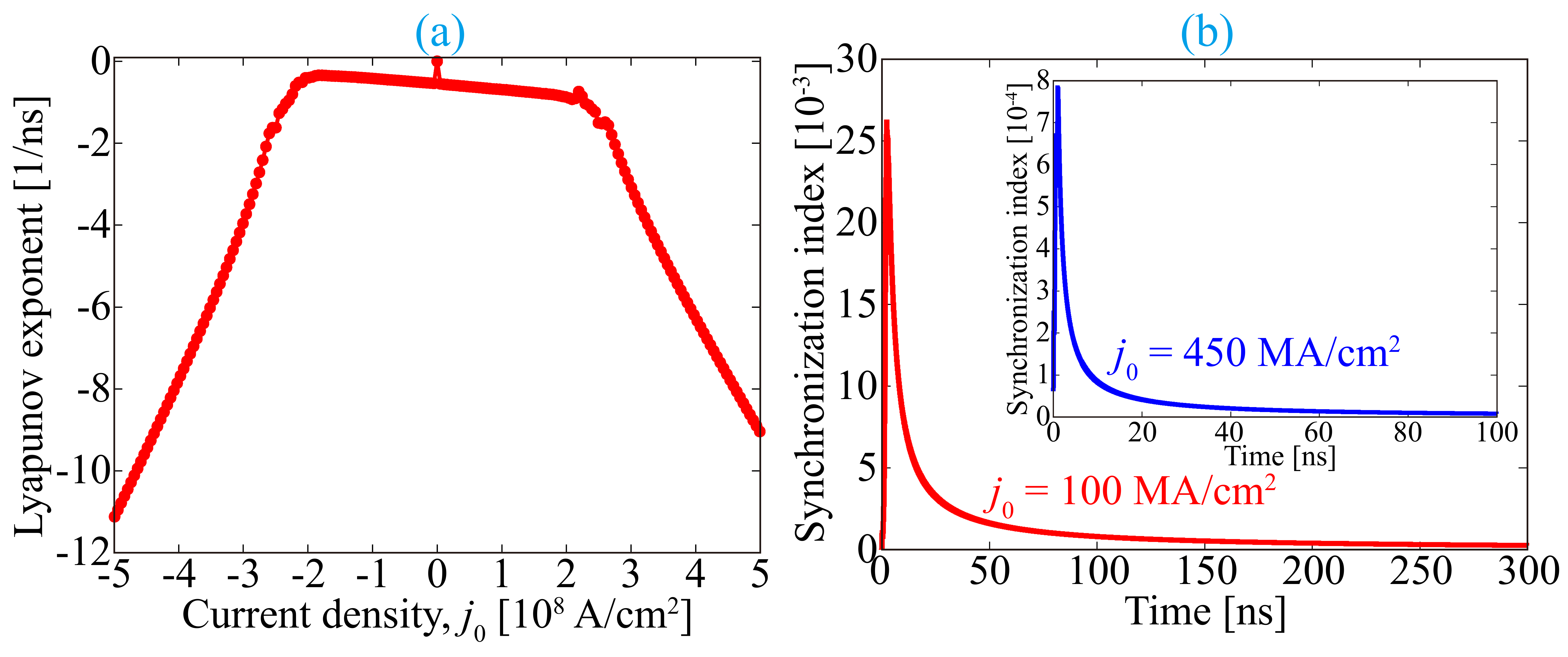}}
\caption{
            (a) Dependence of Lyapunov exponent on current density for an STO with a single free layer. 
            (b) Time evolution of synchronization index at current density of $100$ MA/cm${}^{2}$.
                 The inset shows that for $450$ MA/cm${}^{2}$. 
         \vspace{-3ex}}
\label{fig:fig4}
\end{figure}


\subsection{Lyapunov exponent}
\label{sec:Lyapunov exponent}

Let us study the Lyapunov exponent of the present STO (see also Appendix \ref{sec:AppendixB}). 
The Lyapunov exponent is defined as 
\begin{equation}
  \varLambda
  =
  \lim_{N_{\varLambda}\to \infty}
  \frac{1}{N_{\varLambda}}
  \sum_{i=1}^{N_{\varLambda}}
  \frac{1}{\Delta t}
  \ln
  \frac{\mathscr{D}(t_{i})}{\epsilon}, 
  \label{eq:Lyapunov_exponent_def}
\end{equation}
where $\Delta t$ is the time increment of the LLG equation. 
Here, $\epsilon$ is the distance between two solutions of the LLG equation at every time step, while $\mathscr{D}$ is the distance after the time increment ($\Delta t$) passes. 
We will use the relative angle of the two solutions as the distance $\mathscr{D}(t)$; see Appendix \ref{sec:AppendixB}. 
Note that $\mathscr{D}/\epsilon$ is the expansion rate of the distance $\epsilon$, while the Lyapunov exponent describes an exponential evolution of the distance between the two solutions, $\mathscr{D} \sim \epsilon e^{\varLambda t}$. 

The dependence of the Lyapunov exponent on the current is summarized in Fig. \ref{fig:fig4}(a). 
The exponent is negative throughout the entire current region because the magnetization moves to a fixed point. 
Note that the past information can only be recognized when the output of the system changes with respect to the input signal; if the physical system does not react to an input signal and thus, the output signal is constant, we cannot identify the input signal from the output signal. 
Thus, the short-term memory capacity is often large when the magnitude (absolute value) of the Lyapunov exponent is small. 
This is because a small Lyapunov exponent corresponds to a slow saturation to a fixed point, wherein the history of the input signal is well reflected in the dynamics. 
When the Lyapunov exponent is a large negative value, the magnetization immediately saturates to a fixed point, and the output signal immediately saturates to a constant. 
Therefore, the short-term memory capacity is small when the exponent is a large negative value.  
The Lyapunov exponent is zero when the current density $j_{0}$ is zero because, as mentioned above, the input signal does not drive any dynamics in this case, and thus, the magnetization stays at a fixed point. 

As can be seen from the above, there is a correspondence between the computational capability, quantified by the short-term memory capacity, and the dynamical state, characterized by the Lyapunov exponent. 
However, since the Lyapunov exponent in this case is only negative, the above results reveal only part of the correspondence. 
Note that, according to a mathematical principle, namely the Poincar\'e-Bendixon theorem, the Lyapunov exponent of an STO with a single free layer is negative or zero \cite{bertotti09}. 
Adding an another free layer breaks this mathematical restriction and leads to an appearance of chaos. 
In so doing, a correspondence between the computational capability and the dynamical state of the magnetization will appear. 


\subsection{Synchronization index}
\label{sec:Synchronization index}

Here, let us study the synchronization index of the STO with a single free layer (see also Appendix \ref{sec:AppendixC}). 
The synchronization index is defined as 
\begin{equation}
  \mathcal{S}
  =
  \lim_{N_{\rm s} \to \infty}
  \frac{1}{N_{\rm s}}
  \sum_{i=1}^{N_{\rm s}}
  \mathcal{D}(t_{i}), 
  \label{eq:synchronization_index_def}
\end{equation}
where $N_{\rm s}$ is the number of samples used to evaluate the long-time average of the distance $\mathcal{D}(t)$ between two solutions of Eq. (\ref{eq:LLG_single_free}) with slightly different initial conditions. 

Note that the distance $\mathcal{D}$ here is different from $\mathscr{D}$ used in Sec. \ref{sec:Lyapunov exponent} in the following sense. 
In the evaluation of the Lyapunov exponent, a perturbation with magnitude $\epsilon$ is incremented at every time step by $\Delta t$; see Appendix \ref{sec:AppendixB}. 
On the other hand, in the evaluation of the synchronization index, a perturbation is added only to the initial sate; see Appendix \ref{sec:AppendixC}. 

We should also note that the distance $\mathcal{D}$ here is defined in terms of the dynamical variable used as the output signal; i.e., $\mathcal{D}=|m_{x}^{(1)}-m_{x}^{(2)}|$, where $m_{x}^{(1)}$ and $m_{x}^{(2)}$ are the solutions of $m_{x}$ obtained from the LLG equation, Eq. (\ref{eq:LLG_single_free}), with slightly different initial conditions. 
In Secs. \ref{sec:Short-term memory capacity, Lyapunov exponent, and synchronization index of STO with two free layers} and \ref{sec:Short-term memory capacity, Lyapunov exponent, and synchronization index of STO with two free layers and one reference layer}, $\mathcal{D}$ will be defined in different ways because the output signals from the different STOs depend on different variables. 
The definition of the synchronization index is different from that in the previous paper \cite{akashi20}. 
For example in Ref. \cite{akashi20}, the distance $\mathcal{D}$ is measured in the whole phase space, and is not evaluated from $m_{x}$ only.  
The reason why we define the synchronization differently from the previous work relates to the fact that not all the variables contribute to the output signal used for computing, as mentioned in Sec. \ref{sec:Edges of chaos and echo state property}. 
The difference in the definition of $\mathcal{D}$ is not important in this section; but it will be important in Sec. \ref{sec:Short-term memory capacity, Lyapunov exponent, and synchronization index of STO with two free layers}; see also Appendix \ref{sec:AppendixC}. 

Figure \ref{fig:fig4}(b) shows the time evolution of the temporal synchronization index $\mathcal{S}_{N_{\rm s}}=(1/N_{\rm s})\sum_{i=1}^{N_{\rm s}}\mathcal{D}(t_{i})$ for a low current density, $j_{0}=100$ MA/cm${}^{2}$, where $\mathcal{S}$ in Eq. (\ref{eq:synchronization_index_def}) corresponds to $\lim_{N_{\rm s} \to \infty}\mathcal{S}_{N_{\rm s}}$. 
The synchronization index tends to zero as time increases, as expected from the dynamics shown in Fig. \ref{fig:fig2}, where the magnetization saturates to a fixed point. 
Saturation to zero is also observed for a large current density of $450$ MA/cm${}^{2}$, as shown in the inset of Fig. \ref{fig:fig4}(b). 
We observe similar behavior for the other current density, and find that the synchronization index is zero over a wide range of current density (not shown). 
These results indicate that the output signal ($\propto m_{x}$) eventually becomes independent of its initial state and the STO has the echo state property. 


\section{STO consisting of two free layers}
\label{sec:STO consisting of two free layers}

Now let us examine the STO shown in Fig. \ref{fig:fig1}(b).
We will show that, unlike the results in Sec. \ref{sec:STO with single free layer}, chaos appears in some parameter regions. 


\subsection{LLG equation of STO with two free layers}
\label{sec:LLG equation of STO with two free layers}

The STO consists of two ferromagnets, F${}_{1}$ and F${}_{2}$, separated by a nonmagnetic spacer. 
The LLG equation of the magnetization $\mathbf{m}_{i}$ ($i=1,2$) in F${}_{i}$ layer is given by 
\begin{equation}
  \frac{d \mathbf{m}_{i}}{dt}
  =
  -\gamma
  \mathbf{m}_{i}
  \times
  \mathbf{H}_{i}
  -
  \gamma
  H_{{\rm s}i}
  \mathbf{m}_{i}
  \times
  \left(
    \mathbf{m}_{2}
    \times
    \mathbf{m}_{1}
  \right)
  +
  \alpha_{i}
  \mathbf{m}_{i}
  \times
  \frac{d \mathbf{m}_{i}}{d t},
  \label{eq:LLG_two_free}
\end{equation}
where $H_{{\rm s}i}$ is 
\begin{equation}
  H_{{\rm s}i}
  =
  \frac{\hbar \eta_{i} j}{2e(1 + \lambda_{i} \mathbf{m}_{1}\cdot\mathbf{m}_{2}) M_{i}d_{i}}, 
  \label{eq:H_si}
\end{equation}
The magnetic field \cite{taniguchi18JMMM,taniguchi19} 
\begin{equation}
  \mathbf{H}_{i}
  =
  \begin{pmatrix}
    -4\pi M N_{ix} m_{ix} - H_{{\rm d}i} m_{jx} \\
    -4\pi M N_{iy} m_{iy} - H_{{\rm d}i} m_{jy} \\
    H_{\rm appl} - 4\pi M N_{iz} m_{iz} + 2 H_{{\rm d}i} m_{jz}
  \end{pmatrix},
  \label{eq:H_i}
\end{equation}
includes the dipole field ($\propto H_{{\rm d}i}$) from the other ($j=1,2$ and $j \neq i$) layer, where 
\begin{equation}
  H_{{\rm d}i}
  =
  \pi M_{j}
  \left[
    \frac{\frac{d_{i}}{2} + d_{\rm N} + d_{j}}{\sqrt{r^{2} + \left(\frac{d_{i}}{2} + d_{\rm N} + d_{j} \right)^{2}}}
    -
    \frac{\frac{d_{i}}{2} + d_{\rm N}}{\sqrt{r^{2} + \left(\frac{d_{i}}{2} + d_{\rm N} \right)^{2}}}
  \right].
\end{equation}
Here, $d_{\rm N}$ is the thickness of the spacer layer between the two ferromagnets, which is assumed to be $d_{\rm N}=3$ nm.  
The dynamics of the two magnetizations are coupled via spin-transfer torque and the dipole field. 
The output signal of this STO originates from the magnetoresistance effect between the two free layers and is proportional to $\mathbf{m}_{1}\cdot\mathbf{m}_{2}$. 
The simulations reported in the previous studies assume identical material parameters in two free layers \cite{camsari21} whereas an experimental study used materials with different parameters \cite{zhou19}. 
In this study, we vary the magnetization $M_{i}$ and the Gilbert damping constant $\alpha_{i}$.


\begin{figure*}
\centerline{\includegraphics[width=2.0\columnwidth]{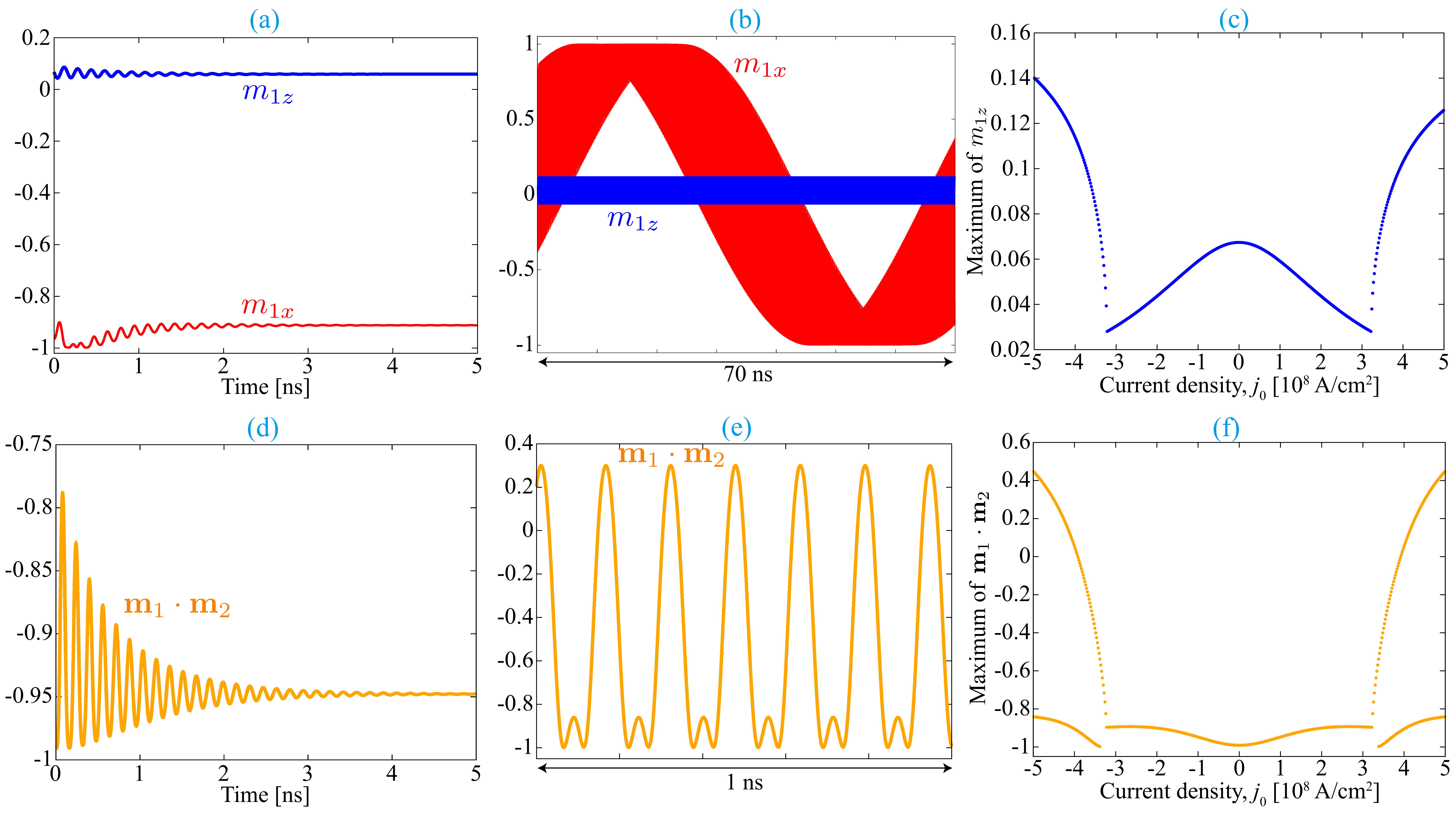}}
\caption{
            Examples of dynamics of the in-plane ($m_{1x}$) and out-of-plane ($m_{1z}$) components of the magnetization for current densities of (a) $100$ and (b) $450$ MA/cm${}^{2}$ in an STO with two free layers. 
            (c) Bifurcation diagram of the local maximum of temporal $m_{1z}$ as a function of current density. 
            Similar data for $\mathbf{m}_{1}\cdot\mathbf{m}_{2}$ are shown in (d), (e), and (f). 
            The values of the parameters in two ferromagnets are identical. 
         \vspace{-3ex}}
\label{fig:fig5}
\end{figure*}



Figure \ref{fig:fig5} shows typical magnetization dynamics of this STO, where two ferromagnets have identical parameters: $M_{1}=M_{2}=1300$ emu/cm${}^{3}$. and $\alpha_{1}=\alpha_{2}=0.01$. 
When the current density is relatively small ($j_{0}=100$ MA/cm${}^{2}$), the magnetizations saturate to a fixed point; the magnetization direction in one (F${}_{1}$) layer is shown in Fig. \ref{fig:fig5}(a). 
On the other hand, when the current density is large ($j_{0}=450$ MA/cm${}^{2}$), an amplitude modulation in the magnetization oscillation occurs; see Fig. \ref{fig:fig5}(b). 
For the discussion later, it will be useful to introduce the bifurcation diagram that summarizes the local maxima of the temporal $m_{1z}$ as a function of the current density; see Fig. \ref{fig:fig5}(c). 
Remember as well that the variable used as the output signal for physical reservoir computing is $\mathbf{m}_{1}\cdot\mathbf{m}_{2}$. 
The time evolution of $\mathbf{m}_{1}\cdot\mathbf{m}_{2}$ for small and large currents and the bifurcation diagram are shown in Figs. \ref{fig:fig5}(d)-\ref{fig:fig5}(f). 
The results indicate that two magnetizations are approximately antiparallel when the current is small, which is due to the fact that the dipole interaction prefers the antiparallel alignment when the magnetizations point in an in-plane direction. 
The spin-transfer torque acting on one ferromagnet also prefers the antiparallel alignment, while that acting on the other prefers the parallel alignment. 
As a result, the magnetization alignment is close to but slighly different from antiparallel, i.e., $\mathbf{m}_{1}\cdot\mathbf{m}_{2}\simeq -1$. 
For a large current, the spin-transfer torque overcomes the damping torque and drives the magnetization oscillations, where $\mathbf{m}_{1}\cdot\mathbf{m}_{2}$ shows two local maxima. 
We emphasize that these dynamics are not chaotic.


\begin{figure}
\centerline{\includegraphics[width=1.0\columnwidth]{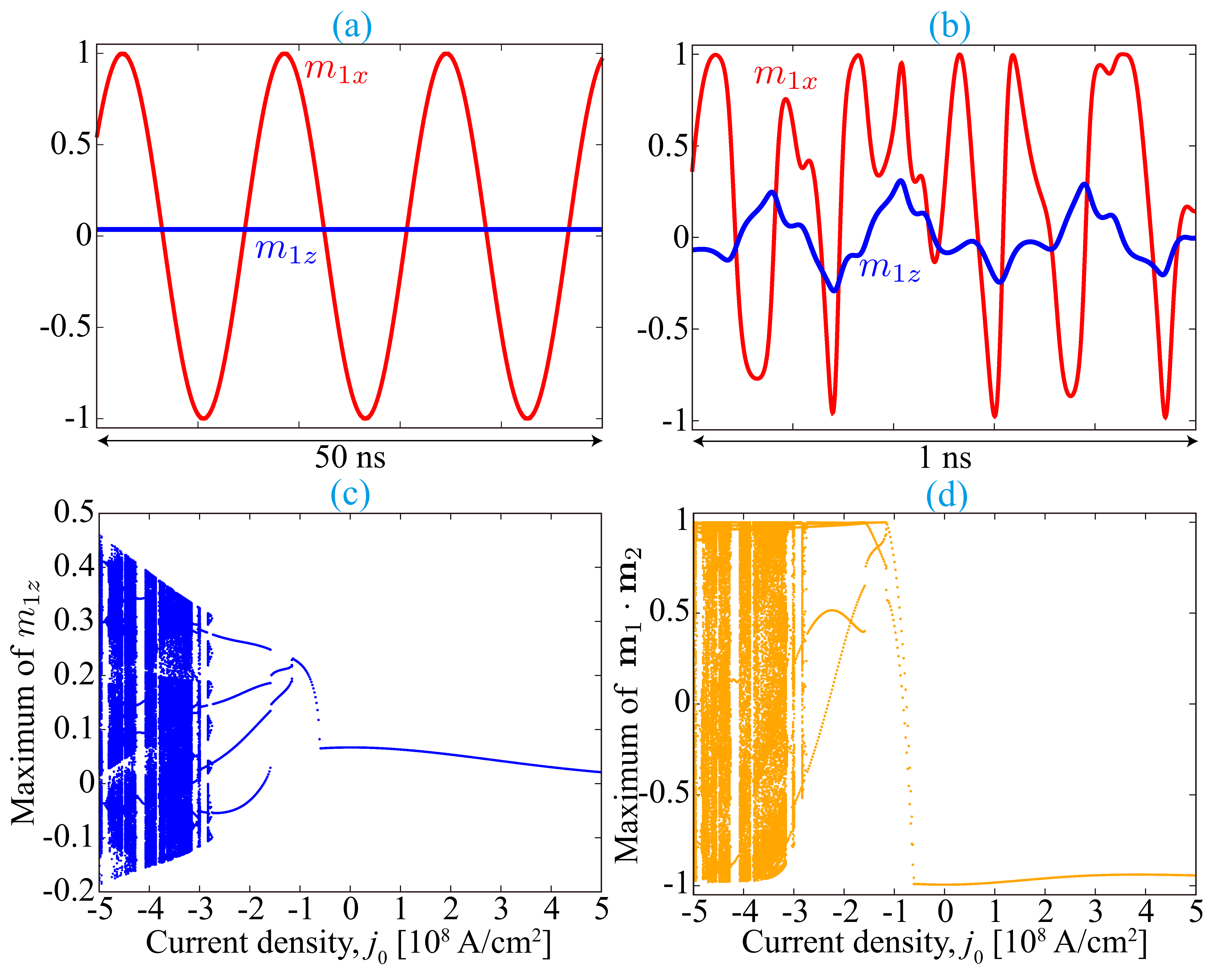}}
\caption{
            Examples of dynamics of the in-plane ($m_{1x}$) and out-of-plane ($m_{1z}$) components of the magnetization for current densities of (a) $100$ and (b) $450$ MA/cm${}^{2}$ in an STO with two free layers.
            Bifurcation diagrams of (c) the local maximum of temporal $m_{1z}$ and (d) $\mathbf{m}_{1}\cdot\mathbf{m}_{2}$ as a function of current density. 
            The values of the magnetization are different as $M_{1}=1300$ emu/cm${}^{3}$ and $M_{2}=2200$ emu/cm${}^{3}$. 
         \vspace{-3ex}}
\label{fig:fig6}
\end{figure}



When the parameters of the two ferromagnets are different, the dynamics become complex \cite{taniguchi19,taniguchi20}. 
As an example, let us suppose that $M_{1}=1300$ emu/cm${}^{3}$, $M_{2}=2200$ emu/cm${}^{3}$, and $\alpha_{1}=\alpha_{2}=0.01$ and study the resulting dynamics. 
In this case, for a positive current, a simple oscillation of the magnetization is excited, as shown in Fig. \ref{fig:fig6}(a) for a current density of $100$ MA/cm${}^{2}$. 
On the other hand, when the current is negative, the dynamics are complex, as shown in Fig. \ref{fig:fig6}(b) for a current density of $450$ MA/cm${}^{2}$. 
The origin of the asymmetry in the dynamics with respect to the current direction is as follows. 
As mentioned, the dipole interaction prefers the antiparallel alignment of magnetizations. 
When the current is positive, the spin-transfer torque acting on the F${}_{1}$ layer also prefers the antiparallel alignment, while that acting on the F${}_{2}$ layer prefers the parallel alignment. 
Note that the strength of the spin-transfer torque acting on the F${}_{2}$ layer is relatively small because it is inversely proportional to the saturation magnetization, and the saturation magnetization $M_{2}$ in the F${}_{2}$ is large in the present case. 
Accordingly, both the dipole interaction and the spin-transfer torques mainly prefer the antiparallel alignment, and the dynamics become relatively simple. 
On the other hand, when the current is negative, the spin-transfer torque acting on the F${}_{1}$ layer prefers the parallel alignment, while that acting on the F${}_{2}$ layer, which is small due to the large $M_{2}$, prefers the antiparallel alignment. 
Thus, while the dipole interaction prefers the antiparallel alignment, the spin-transfer torques mainly prefer the parallel alignment. 
As a result of competition between them, the dynamics become complex; in fact, as clarified from the Lyapunov exponent shown below, the dynamics in Fig. \ref{fig:fig6}(b) can be classified to chaos. 
The complexity of the dynamics can be seen in the bifurcation diagrams of $m_{1z}$ and $\mathbf{m}_{1}\cdot\mathbf{m}_{2}$ [Figs. \ref{fig:fig6}(c) and \ref{fig:fig6}(d)]. 
The broad distributions, as well as the window structures, imply the appearance of chaos in the negative current region \cite{strogatz01}.


\begin{figure*}
\centerline{\includegraphics[width=2.0\columnwidth]{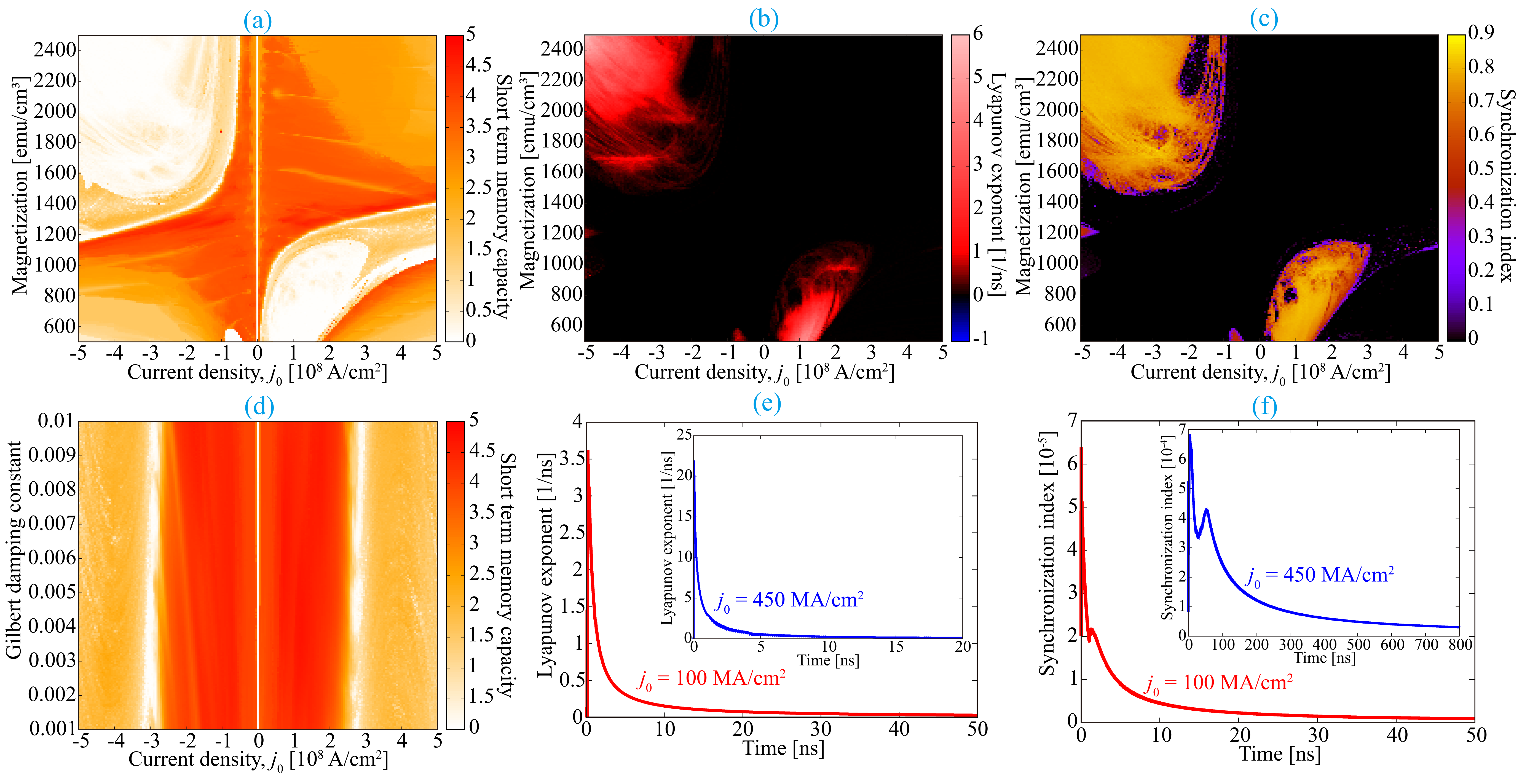}}
\caption{
            (a) Short-term memory capacity, (b) Lyapunov exponent, and (c) synchronization index of STO with two free layers, where the horizontal axis is the current density and the vertical axis is the saturation magnetization in F${}_{2}$ layer. 
            (d) Short-term memory where the vertical axis the Gilbert damping constant of the F${}_{2}$ layer. 
            Examples of time evolution of the temporal (e) Lyapunov exponent and (f) synchronization index for current density of $100$ MA/cm${}^{2}$. 
            The insets show those for the current density of $450$ MA/cm${}^{2}$. 
            The Gilbert damping constant of the F${}_{2}$ layer is $0.01$. 
         \vspace{-3ex}}
\label{fig:fig7}
\end{figure*}



\subsection{Short-term memory capacity, Lyapunov exponent, and synchronization index of STO with two free layers}
\label{sec:Short-term memory capacity, Lyapunov exponent, and synchronization index of STO with two free layers}

Figures \ref{fig:fig7}(a)-\ref{fig:fig7}(c) show the short-term memory capacity, the Lyapunov exponent, and the synchronization index of the STO, where the magnetization $M_{2}$ in the F${}_{2}$ layer is plotted on the vertical axis. 
The definition of the distance $\mathcal{D}$ in the case of the synchronization index is the difference in the values of $\mathbf{m}_{1}\cdot\mathbf{m}_{2}$ calculated under slightly different initial conditions. 
The results in the figures indicate the followings. 
First, the short-term memory capacity is almost zero when the system is chaotic, i.e., when the Lyapunov exponent is positive. 
Second, in addition to the boundary between zero and finite short-term memory capacity, there are boundaries along which an enhancement in the capacity can be observed. 
We consider that such an enhancement appears at the edge of the echo state property, as described below. 
Third, the short-term memory capacity is larger than that of the STO with a single free layer; for example, it is $4.60$ when $j_{0}=344$ MA/cm${}^{2}$ and $M_{2}=1380$ emu/cm${}^{3}$. 


We should note that the Lyapunov exponent of the present STO is at least zero, or positive (see also Appendix \ref{sec:AppendixB}). 
This is due to the axial symmetry around the $z$-axis, where rotations of the two magnetizations around the $z$ axis through the same angle do not change the system energy. 
As a result, a perturbation given to the phases of the magnetizations in the $xy$ plane remains finite. 
This means that the maximum Lyapunov exponent is at least zero. 
Therefore, one might consider that the present STO does not have the echo state property. 
However, the rotations of $\mathbf{m}_{1}$ and $\mathbf{m}_{2}$ around the $z$-axis through the same angle do not change the output signal of the STO, $\mathbf{m}_{1}\cdot\mathbf{m}_{2}$. 
In this sense, the output signal has the echo state property in some parameter regions, even where the (maximum) Lyapunov exponent is zero. 
Therefore, to reveal the echo state property of the output signal, we evaluated the synchronization index of $\mathbf{m}_{1}\cdot\mathbf{m}_{2}$, in addition to the Lyapunov exponent. 
Here, we found that the short-term memory capacity is finite and can be large at the edge of the echo state property; compare Figs. \ref{fig:fig7}(a) and \ref{fig:fig7}(c). 


Now let us examine the effect on the short-term memory capacity of varying the damping constant $\alpha_{2}$ in the F${}_{2}$ layer for saturation magnetizations of $M_{1}=M_{2}=1300$ emu/cm${}^{3}$; see Fig. \ref{fig:fig7}(d). 
In this case, the short-term memory capacity is large when the current is small. 
Also, chaos is absent in this parameter region; see Fig. \ref{fig:fig7}(e), where the temporal Lyapunov exponents for small and large current densities tend to be zero. 
The fact that the short-term memory capacity remains finite also implies the presence of the echo state property; see also Fig. \ref{fig:fig7}(f), where the temporal synchronization indexes also saturate to zero. 


\section{STO consisting of two free and one reference layers}
\label{sec:STO consisting of two free and one reference layers}

In this section, we the STO schematically shown in Fig. \ref{fig:fig1}(c). 
The spin-transfer torque from the reference layer provides an additional torque and change the dynamical state and the computational capability from those of the STOt studied in Sec. \ref{sec:STO consisting of two free layers}.  

\subsection{LLG equation of STO with two free and one reference layers}
\label{sec:LLG equation of STO with two free and one reference layers}

The LLG equations of the magnetizations in the F${}_{1}$ and F${}_{2}$ layers are given by 
\begin{equation}
\begin{split}
  \frac{d \mathbf{m}_{1}}{dt}
  =&
  -\gamma
  \mathbf{m}_{1}
  \times
  \mathbf{H}_{1}
  -
  \gamma
  H_{{\rm s}1}
  \mathbf{m}_{1}
  \times
  \left(
    \mathbf{m}_{2}
    \times
    \mathbf{m}_{1}
  \right)
\\
  &+
  \gamma
  H_{\rm s}
  \mathbf{m}_{1}
  \times
  \left(
    \mathbf{p}
    \times
    \mathbf{m}_{1}
  \right)
  +
  \alpha_{1}
  \mathbf{m}_{1}
  \times
  \frac{d \mathbf{m}_{1}}{d t},
  \label{eq:LLG_1}
\end{split}
\end{equation}
\begin{equation}
  \frac{d \mathbf{m}_{2}}{dt}
  =
  -\gamma
  \mathbf{m}_{2}
  \times
  \mathbf{H}_{2}
  -
  \gamma
  H_{{\rm s}2}
  \mathbf{m}_{2}
  \times
  \left(
    \mathbf{m}_{2}
    \times
    \mathbf{m}_{1}
  \right)
  +
  \alpha_{2}
  \mathbf{m}_{2}
  \times
  \frac{d \mathbf{m}_{2}}{d t}.
  \label{eq:LLG_2}
\end{equation}
Assuming that the total output signal is dominated by the magnetoresistance effect between the reference and F${}_{1}$ layer, the output signal is proportional to $\mathbf{m}_{1}\cdot\mathbf{p}=m_{1x}$.


\begin{figure}
\centerline{\includegraphics[width=1.0\columnwidth]{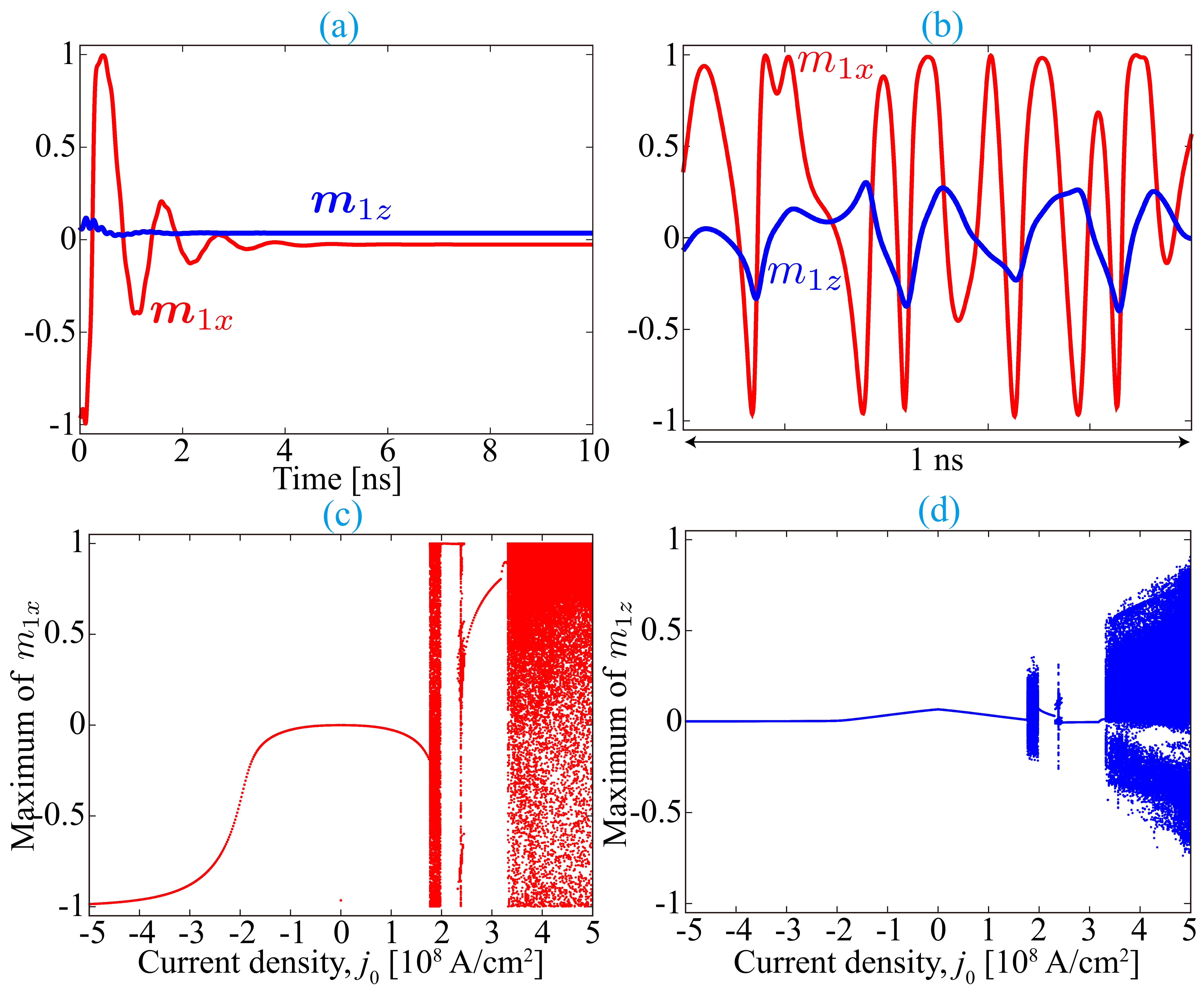}}
\caption{
            Examples of dynamics of the in-plane ($m_{1x}$) and out-of-plane ($m_{1z}$) components of the magnetization for current densities of (a) $100$ and (b) $450$ MA/cm${}^{2}$ in STO with two free layers and one reference layer.
            Bifurcation diagrams of the local maxima of temporal (c) $m_{1x}$ and (d) $m_{1z}$ as a function of current density. 
         \vspace{-3ex}}
\label{fig:fig8}
\end{figure}


Figures \ref{fig:fig8}(a) and \ref{fig:fig8}(b) show typical dynamics excited in the STO for small ($100$ MA/cm${}^{2}$) and large ($450$ MA/cm${}^{2}$) currents. 
The two ferromagnets have identical parameters: $M_{1}=M_{2}=1300$ emu/cm${}^{3}$ and $\alpha_{1}=\alpha_{2}=0.01$. 
When the current is small, the magnetizations saturate to a fixed point. 
Unlike the STO studied in Sec. \ref{sec:STO consisting of two free layers}, complex dynamics appear for a large current, even when the parameters of the two ferromagnets are identical, due to the spin-transfer torque from the reference layer acting on only the F${}_{1}$ layer. 
The bifurcation diagrams of $m_{1x}$ and $m_{1z}$ [Figs. \ref{fig:fig8}(c) and \ref{fig:fig8}(d)] show that complex structures appear in the positive current region. 
These results imply chaos in the positive current region. 

The asymmetry of the dynamics with respect to the current direction arises for the following reason. 
First, let us consider the negative current case. 
The spin-transfer torque from the reference layer acting on the F${}_{1}$ layer moves $\mathbf{m}_{1}$ in the $-x$ direction. 
Then, $\mathbf{m}_{2}$ moves in the $+x$ direction to minimize the dipole interaction energy. 
The spin-transfer torque from the F${}_{2}$ acting on the F${}_{1}$ layer prefers the parallel alignment of the magnetizations, and thus, tries to move $\mathbf{m}_{1}$ in the $+x$ direction. 
However, this motion is compensated against with the spin-transfer torque from the reference layer, and $\mathbf{m}_{1}$ remains in the $-x$ direction. 
The spin-transfer torque from the F${}_{1}$ acting on the F${}_{2}$ layer prefers the antiparallel alignment of the magnetization, and thus, $\mathbf{m}_{2}$ also remains in the $+x$ direction.
Accordingly, the two magnetizations stay the fixed points. 

Next, let us consider the  positive current case. 
The spin-transfer torque from the reference layer acting on the F${}_{1}$ layer moves $\mathbf{m}_{1}$ in the $+x$ direction. 
Accordingly, $\mathbf{m}_{1}$ and $\mathbf{m}_{2}$ try to point in the $+x$ and $-x$ direction, respectively. 
However, the spin-transfer torque from the F${}_{1}$ acting on the F${}_{2}$ layer in this case prefers the parallel alignment of the magnetizations, and thus, $\mathbf{m}_{2}$ cannot remain in the $-x$ direction. 
As a result, the magnetizations do not saturate to a fixed point when the current magnitude is large.


\begin{figure*}
\centerline{\includegraphics[width=2.0\columnwidth]{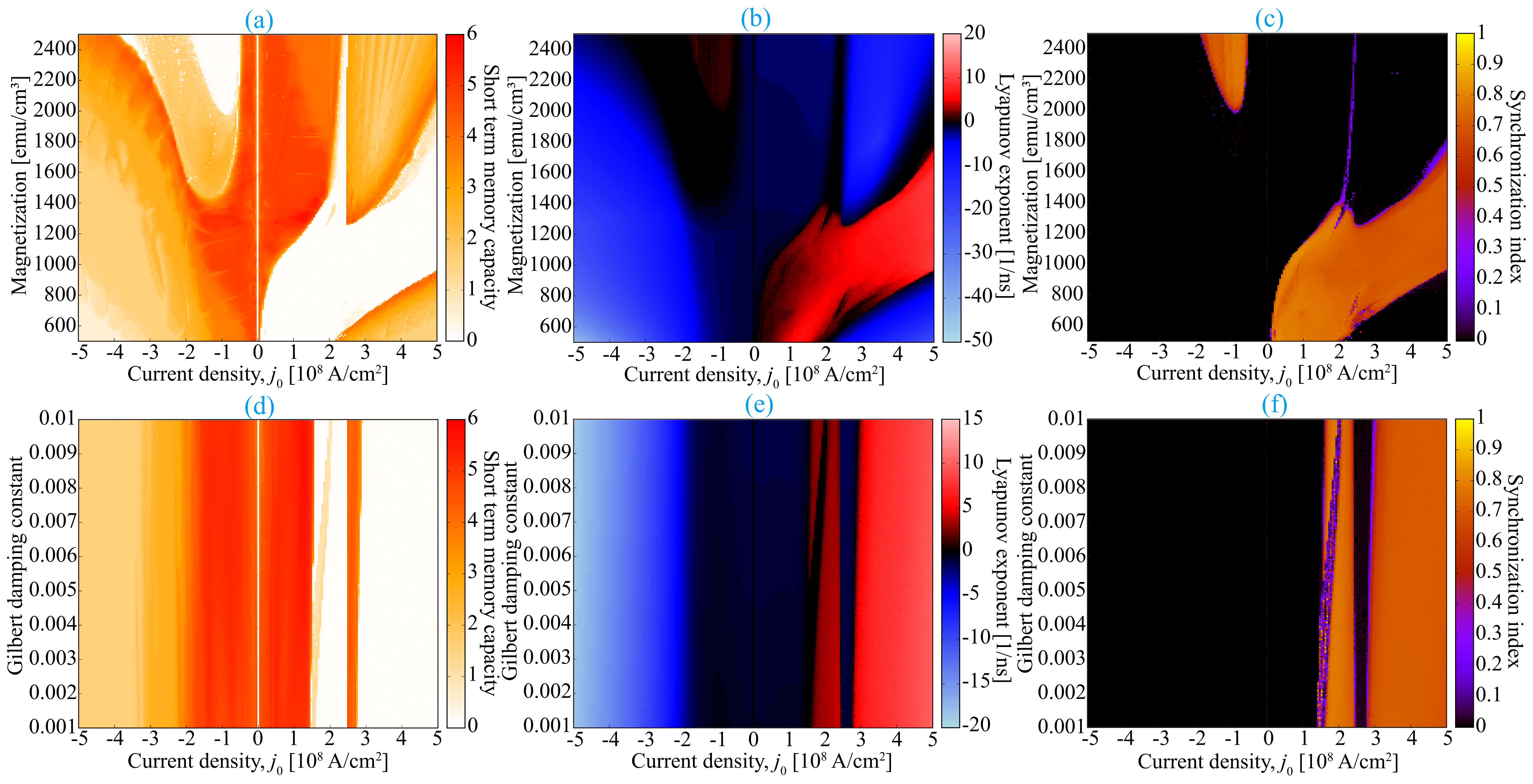}}
\caption{
            (a) Short-term memory capacity, (b) Lyapunov exponent, and (c) synchronization index of an STO with two free layers and one reference layer, where the horizontal axis is the current density and the vertical axis is the saturation magnetization in the F${}_{2}$ layer. 
            (d), (e), and (f) are the same as (a), (b), and (c) but as a function of the Gilbert damping constant in the F${}_{2}$. 
         \vspace{-3ex}}
\label{fig:fig9}
\end{figure*}



\subsection{Short-term memory capacity, Lyapunov exponent, and synchronization index of STO with two free layers and one reference layer}
\label{sec:Short-term memory capacity, Lyapunov exponent, and synchronization index of STO with two free layers and one reference layer}

Figure \ref{fig:fig9}(a)-\ref{fig:fig9}(c) summarizes the short-term memory capacity, the Lyapunov exponent, and the synchronization index of the present STO, where the saturation magnetization $M_{2}$ in the F${}_{2}$ layer is plotted in the vertical axes. 
The distance $\mathcal{D}$ for the synchronization index is evaluated from $m_{1x}$. 
The maximum short-term memory capacity, $5.72$ for $j_{0}=127$ MA/cm${}^{2}$ and $M_{2}=1320$ emu/cm${}^{3}$, is larger than those of the STO with one free layer and the STO with two free layers. 
These results in these sections, as well as here, indicate that adding another free layer makes the magnetization dynamics complex and helps to enhance the computational capability of STO-based physical reservoir computing. 
The short-term memory capacity again becomes zero when the system is in a chaotic state, where the Lyapunov exponent is positive; see Figs. \ref{fig:fig9}(a) and \ref{fig:fig9}(b). 
The maximum value of the short-term memory capacity appears near the edge of the echo state property. 
Here, the parameter regions corresponding to the zero Lyapunov exponent are relatively limited especially in the positive current region, so the edges of chaos and the echo state property nearly overlap in the present STO. 
This might be due to the spin-transfer torque from the reference layer, which breaks the systems's axial symmetry around the $z$-axis and reduces the parameter region corresponding to the zero Lyapunov exponent, compared to those in Sec. \ref{sec:Short-term memory capacity, Lyapunov exponent, and synchronization index of STO with two free layers}. 
Similar behaviors are found when the damping constant $\alpha_{2}$ in the F${}_{2}$ layer is varied, as shown by the plots of the short-term memory capacity, Lyapunov exponent, and synchronization index in Figs. \ref{fig:fig9}(d)-\ref{fig:fig9}(f). 
The maximum short-term memory capacity is $5.76$ for $j_{0}=127$ MA/cm${}^{2}$ and $\alpha_{2}=0.00847$. 
As shown in Figs. \ref{fig:fig9}(a) and \ref{fig:fig9}(d) that the maximum short-term memory capacity occurs in a relatively low current region, while low computational capability dominates in the relatively high current region due to the appearance of chaos. 
This fact might make the low current region preferable for physical reservoir computing. 

Here, we have shown the dependence of the short-term memory capacity on the parameters in the F${}_{2}$ layer. 
Similar behaviors, such as maximization of the short-term memory capacity near the edge of the echo state property, occur even when the parameters in the F${}_{1}$ layer are varied; see Appendix \ref{sec:AppendixD}. 


\section{Conclusions}
\label{sec:Conclusions}

In summary, we studied the magnetization dynamics in STOs with two free layers. 
It was shown that adding another free layer makes the dynamical output signal complex due to the coupled motion of the magnetizations via the spin-transfer torques and the dipole field. 
For example, in addition to the saturation of the magnetization to a fixed point found in the STO with a single free layer, an amplitude-modulated oscillation was found in the STO consisting of two free layers. 
The STOs with two free layers also showed chaotic dynamics particularly in the large current region. 
These complex dynamics mainly appear when structural asymmetries, such as a difference in parameters and/or the presence of the reference layer, exist. 
We investigated the computational capability of these STOs for physical reservoir computing by evaluating the short-term memory capacity. 
The maximum values for the STOs with two free layers were larger than that of the STO with a single free layer. 
Through the evaluations of the Lyapunov exponent and the synchronization index, it was shown that the short-term memory capacity is maximized near the edge of the echo state property. 
We note that the short-term memory capacity is the linear component of the information processing capacity \cite{dambre12,kubota21}, and the total information processing capacity is bounded by the linearly independent output of the system. 
Therefore, an increase in the short-term memory capacity does not guarantee an increase in the total information processing capacity directly; the nonlinear components of the information processing capacity might be suppressed in STOs with two free layers. 
In future, the total information processing capacity should be analyzed.


\section*{Acknowledgements}

The results were partially obtained from the project 
``Innovative AI Chips and Next-Generation Computing Technology Development/(2) Development of Next-Generation Computing Technologies/Exploration of Neuromorphic Dynamics towards Future Symbiotic Society'' commissioned by NEDO. 
T.Y. is supported by JSPS KAKENHI Grant No. 21K14526. 
T.T. is supported by JSPS KAKENHI Grant No. 20H05655. 


\appendix


\section{Method of evaluating short-term memory capacity}
\label{sec:AppendixA}

Here, we summarize the details of the method of evaluating the short-term memory capacity. 
The short-term memory capacity is a kind of information processing capacity \cite{dambre12,kubota21} and quantifies task-independent computational capability. 
We assume that a series of pulse input signals $r_{k}$ ($k=1,2,\cdots,N_{\rm L}$) is injected into the physical reservoir, where the suffix $k$ distinguish the order of the input signal. 
In the main text, we used a binary input signal $b_{k}=0,1$ as the input signal $r_{\rm k}$. 
Another kind of input signal can be found in, for example, Ref. \cite{kubota21}, where a uniformly distributed random number ($0 \le r_{k} \le 1$ or $-1 \le r_{k} \le 1$) is used. 
We define the target data $z_{k,D}$ from the input signal $r_{k}$. 
Here, $D$ is an integer called the delay ($D=0,1,2,\cdots$). 
An aim of physical reservoir computing is to recognize the past input data from the present output signal, and therefore, it is necessary to introduce a delay to distinguish the past input data. 
For example, in the evaluation of the short-term memory capacity, $z_{k,D}$ is $b_{k-D}$ \cite{fujii17,tsunegi18} (or $r_{k-D}$ \cite{kubota21}); i.e., the target data are the input data injected $D$ times before from the present input signal. 
Another example of $z_{k,D}$ is $z_{k,D}=\sum_{j=0}^{D}b_{k-D+j}$ (mod $2$) for the evaluation of parity-check capacity \cite{fujii17,tsunegi18}. 
The target data of the information processing capacity \cite{dambre12,kubota21} are, in general, nonlinear combination of $r_{k-D}$. 
After defining the target data, we introduce the weight $w_{D,i}$ to minimize 
\begin{equation}
  \sum_{k=1}^{N_{\rm L}}
  \left(
    \sum_{i=1}^{N_{\rm node}+1}
    u_{k,i}
    w_{D,i}
    -
    z_{k,D}
  \right)^{2}, 
 \end{equation}
where the output data from the $i$th (virtual) node in the presence of the $k$th input is denoted as $u_{k,i}$. 
When physical reservoir is a many body system, the suffix $i$ distinguishes each body. 
On the other hand, in the present paper, we use a single STO. 
 In this case, a time-multiplexing method \cite{fujii17} is applied in order to introduce virtual neurons, $u_{k,i}=u[t_{0}+(k-1+i/N_{\rm node})t_{\rm p}]$, where $t_{0}$ is the initial time at which the input signal is injected while $N_{\rm node}$ is the number of virtual neurons. 
 The function $u(t)$ is the output signal from the STO; for example, in the case of the STO with a single free layer studied in Sec. \ref{sec:STO with single free layer} , the experimentally measured quantity is $m_{x}$, and thus, $u(t)=m_{x}(t)$. 
The process determining the weight is called learning. 
The number of the input signal used for learning is $N_{\rm L}$. 
Note that a weight should be introduced for each target data.


\begin{figure*}
\centerline{\includegraphics[width=2.0\columnwidth]{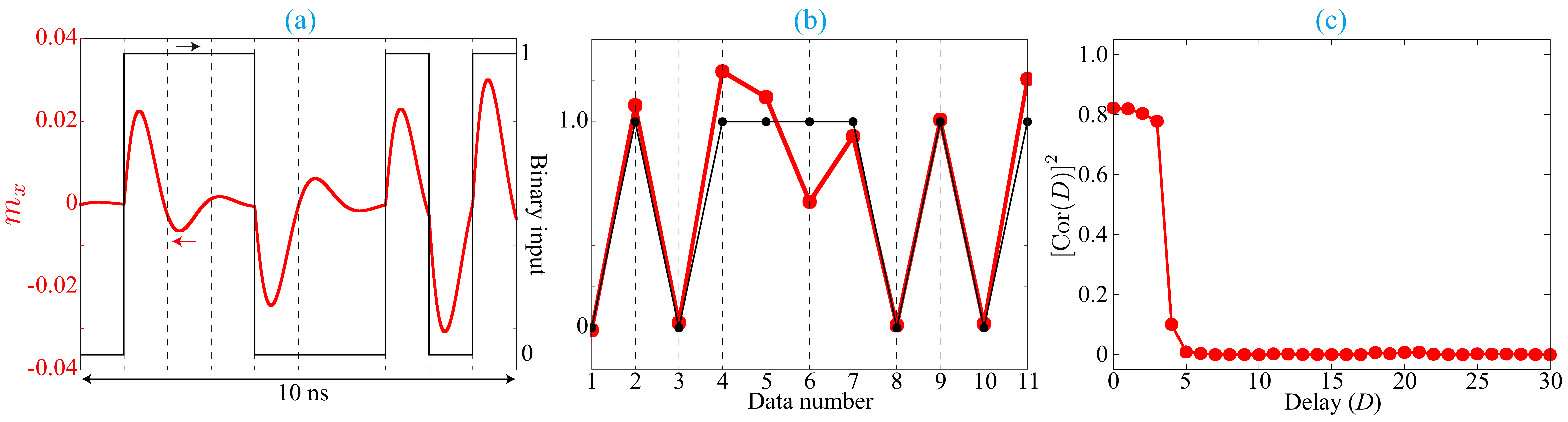}}
\caption{
            (a) Temporal dynamics of $m_{x}$ (red) and random binary input  (black) for a current density of $10$ MA/cm${}^{2}$. 
                 The pulse width is $t_{\rm p}=1$ ns. 
            (b) Examples of target data (black) and system output (red) for $D=1$.  
            (c) Dependence of $[{\rm Cor}(D)]^{2}$ on $D$. 
         \vspace{-3ex}}
\label{fig:fig10}
\end{figure*}


Next, we inject a different series of pulses $r_{n}^{\prime}$ ($n=1,2,\cdots,N_{\rm L}^{\prime}$), where the prime symbol is added to quantities to distinguish them from those used in learning. 
 The number $N_{\rm L}^{\prime}$ of input data is not necessarily the same as the number used in learning, i.e., $N_{\rm L}\neq N_{\rm L}^{\prime}$. 
Then, from the output data $u_{n,i}^{\prime}$, which is the response of the physical reservoir to the injection of $r_{n}^{\prime}$, and using the weight $w_{D,i}$ determined by learning, we define system output as 
\begin{equation}
   y_{n,D}^{\prime}
   =
   \sum_{i=1}^{N_{\rm node}+1}
   u_{n,i}
   w_{D,i}. 
\end{equation}
If the learning is done well, $y_{n,D}^{\prime}$ will reproduce the target data $z_{k,D}^{\prime}$ defined from $r_{n}^{\prime}$. 
To quantify the reproducibility, we can use the correlation coefficient, 
\begin{widetext}
\begin{equation}
  {\rm Cor}(D)
  =
  \frac{\sum_{n=1}^{N_{\rm L}^{\prime}} \left(z_{n,D}^{\prime} - \langle z_{n,D}^{\prime} \rangle \right) \left(y_{n,D}^{\prime} - \langle y_{n,D}^{\prime} \rangle \right)}
     {\sqrt{ \sum_{n=1}^{N_{\rm L}^{\prime}} \left(z_{n,D}^{\prime} - \langle z_{n,D}^{\prime} \rangle \right)^{2} \sum_{n=1}^{N_{\rm L}^{\prime}} \left(y_{n,D}^{\prime} - \langle y_{n,D}^{\prime} \rangle \right)^{2} }}.
  \label{eq:cor}
\end{equation}
\end{widetext}
The component-wise information processing capacity is defined as 
\begin{equation}
   C(z_{n,D}^{\prime})
   =
   \left[
     {\rm Cor}(D)
   \right]^{2}.
\end{equation}
The magnitude of the correlation coefficient is unity when the system output $y_{n,D}^{\prime}$ completely reproduces the target data $z_{n,D}^{\prime}$. 
On the other hand, the correlation coefficient is zero when the input signal cannot reproduce the input data. 
Therefore, the component-wise information processing capacity quantifies the reproducibility of the target data. 
Note that the component-wise information processing capacity is introduced for each target data $z_{n,D}^{\prime}$, and is independent of the suffix $n$ because the average with respect to the input pulse is calculated in Eq. (\ref{eq:cor}). 
For example, Ref. \cite{kubota21} evaluates the component-wise information processing capacity of several physical reservoirs, where the capacities are distinguished by the nonlinearity of the target data and the delay $D$. 
In the evaluation of the short-term memory capacity, we restrict the target data to being a linear combination of the input data, i.e., $z_{k,D}=b_{k-D}$, and define the short-term memory capacity as the sum of the component-wise information processing capacity with respect to the delay $D$, i.e., 
\begin{equation}
  C_{\rm STM}
  =
  \sum_{D=1}^{D_{\rm max}}
  C(b_{n-D}^{\prime}); 
  \label{eq:C_STM_def}
\end{equation}
see also, for example, Ref. \cite{tsunegi18,jaeger02} for the definition of the short-term memory capacity. 
In many cases \cite{tsunegi18}, the correlation coefficient ${\rm Cor}(D)$ becomes negligibly small for a large delay when the physical reservoir has the echo state property, and thus, past information fades in time. 
In such a case, the value of the information processing capacity will be independent of the maximum delay $D_{\rm max}$. 
As mentioned, the component-wise capacity quantifies the reproducibility of the target data, and the target data for the evaluation of the short-term memory capacity is the input data as is.
Therefore, the short-term memory capacity can be, roughly speaking, regarded as the number of the past input data reproduced from the present output, as mentioned in the main text. 
Note that, in some cases \cite{akashi20}, $C(z_{n,D})$ for $D=0$ is included in the definition of the capacity. 
In the present study, we use $N_{\rm L}=N_{\rm L}^{\prime}=1000$ random binary data, $N_{\rm node}=250$ nodes, and $D_{\rm max}=30$. 


Figure \ref{fig:fig10} summarizes examples of these procedures for an STO with single free layer with the current density of $10$ MA/cm${}^{2}$. 
In Fig. \ref{fig:fig10}(a), an example of a series of random binary input signal and the dynamics of $m_{x}$ are shown. 
The value of $m_{x}$ changes with respect to the input signal; from this dynamical response, the input signal can be identified. 
Figure \ref{fig:fig10}(b) shows an example of the reproduction of the input data with delay $D=1$. 
As can be seen, the reproducibility decreases with the delay increasing. 
Figure \ref{fig:fig10}(c) shows the dependence of $[{\rm Cor}(D)]^{2}$ on the delay $D$. 
The short-term memory capacity is obtained as a sum of these $[{\rm Cor}(D)]^{2}$.


In the determination of the weight, the output $u_{k,i}$ should be independent of the initial state. 
This is because the initial state of physical system is often uncontrollable, and the computational capability should not depend on such uncontrollable variables. 
In addition, the output $u_{n,i}^{\prime}$ used in the evaluation of the capacity should be independent of the input data $r_{k,i}$ used in the determination of weight. 
This is because there should be no correlation between the learning and the evaluation of capacity. 
As mentioned in the main text, if the physical reservoir has echo state property, the dynamical state will be independent of the past state by injecting random input signal. 
Therefore, before the determination of the weight, we inject $300$ random binary input signals for STO. 
Similarly, after determining the weight and before evaluating the capacity, we also inject different $300$ random binary input signals to erase a correlation between the learning and the evaluation.
These processes are called washout. 
As can be seen from these examples, the echo state property is a necessary factor for physical reservoir computing. 


\section{Method of evaluating Lyapunov exponent}
\label{sec:AppendixB}

Here, let us summarize the method of evaluating the Lyapunov exponent. 
For simplicity, we will use an STO with a single free layer as an example, for a while. 

We denote the solution of the LLG equation with a certain initial condition as $\mathbf{m}(t)$. 
At a certain time $t_{0}$, we introduce $\mathbf{m}^{(1)}(t_{0})$, which points in a slightly different direction from $\mathbf{m}(t_{0})$ with distance $\epsilon$. 
We emphasize that there is no correlation between $\mathbf{m}(t)$ and $\mathbf{m}^{(1)}(t_{0})$. 
The distance is the relative angle of two magnetizations, i.e., $\epsilon=\cos^{-1}\left[ \mathbf{m}(t_{0}) \cdot \mathbf{m}^{(1)}(t_{0}) \right]$. 
Solving the LLG equations for $\mathbf{m}(t_{0})$ and $\mathbf{m}^{(1)}(t_{0})$, we obtain $\mathbf{m}(t_{0}+\Delta t)$ and $\mathbf{m}^{(1)}(t_{0}+\Delta t)$.  
Then, we define a temporal Lyapunov exponent at time $t_{1}=t_{0}+\Delta t$ as 
\begin{equation}
  \varLambda(t_{1})
  =
  \frac{1}{\Delta t}
  \ln
  \frac{\mathscr{D}(t_{1})}{\epsilon},
\end{equation}
where $\mathscr{D}(t_{1})=\cos^{-1}[\mathbf{m}(t_{0}+\Delta t)\cdot\mathbf{m}^{(1)}(t_{0}+\Delta t)]$ is the distance between $\mathbf{m}(t_{0}+\Delta t)$ and $\mathbf{m}^{(1)}(t_{0}+\Delta t)$. 
Next, we introduce $\mathbf{m}^{(2)}(t_{0}+\Delta t)$ by moving $\mathbf{m}(t_{0}+\Delta t)$ in the direction of $\mathbf{m}^{(1)}(t_{0}+\Delta t)$ through the distance $\epsilon$. 
Solving the LLG equations of $\mathbf{m}(t_{0}+\Delta t)$ and $\mathbf{m}^{(2)}(t_{0}+\Delta t)$ yields $\mathbf{m}(t_{0}+2\Delta t)$ and $\mathbf{m}^{(2)}(t_{0}+2 \Delta t)$. 
Then, the temporal Lyapunov exponent at time $t_{2}=t_{0}+2\Delta t$ is defined as $\varLambda(t_{2})=(1/\Delta t)\ln[\mathscr{D}(t_{2})/\epsilon]$, where $\mathscr{D}(t_{2})$ is the distance between $\mathbf{m}(t_{2})$ and $\mathbf{m}^{(2)}(t_{2})$. 

Now let us generalize the above procedure. 
At $t_{n}=t_{0}+n \Delta t$, we introduce $\mathbf{m}^{(n+1)}(t_{n})$ by moving $\mathbf{m}(t_{n})$ in the direction of $\mathbf{m}^{(n)}(t_{n})$ through a fixed distance $\epsilon$. 
Solving the LLG equation, we obtain $\mathbf{m}(t_{n+1})$ and $\mathbf{m}^{(n+1)}(t_{n+1})$. 
From the distance $\mathscr{D}(t_{n+1})=\cos^{-1}[\mathbf{m}(t_{n+1})\cdot\mathbf{m}^{(n+1)}(t_{n+1})]$ between $\mathbf{m}(t_{n+1})$ and $\mathbf{m}^{(n+1)}(t_{n+1})$, the temporal Lyapunov exponent at $t=t_{n+1}$ is defined as $\varLambda(t_{n+1})=(1/\Delta t)\ln[\mathscr{D}(t_{n+1})/\epsilon]$. 
Then, the Lyapunov exponent is defined as
\begin{equation}
  \varLambda
  =
  \lim_{N_{\varLambda}\to\infty}
  \frac{1}{N_{\varLambda}}
  \sum_{i=1}^{N_{\varLambda}}
  \varLambda(t_{i}). 
\end{equation}
For STOs including two free layers, we should define $\mathbf{m}_{i}^{(n+1)}(t_{n})$ to make the total distance, i.e., the sum of the distances between $\mathbf{m}_{i}(t_{n})$ and $\mathbf{m}_{i}^{(n+1)}(t_{n})$, equal to $\epsilon$; see also Ref. \cite{taniguchi19PRB}, where a similar method for an STO with a feedback circuit is developed. 

As can be seen from this explanation, the distance between two samples is given at every time step $t_{n}$, contrary to the evaluation of the synchronization index explained in Appendix \ref{sec:AppendixC} below, where the perturbation is given at the initial time only. 

Note that the difference $\mathbf{m}^{(n+1)}(t_{n})-\mathbf{m}(t_{n})$ between $\mathbf{m}(t_{n})$ and $\mathbf{m}^{(n+1)}(t_{n})$ corresponds to the direction along which the difference expands the most. 
Therefore, the Lyapunov exponent estimated above is the maximum (or largest) Lyapunov exponent, which quantifies the maximum expansion rate from the initial difference. 
There are $n$ Lyapunov exponent, $\varLambda_{1}$, $\varLambda_{2}$, $\cdots$, $\varLambda_{n}$ ($\varLambda_{1} \ge \varLambda_{2} \ge \cdots \ge \varLambda_{n}$) for a system with $n$ dimensions, and $\varLambda$ above corresponds to $\varLambda_{1}$. 
Ather Lyapunov exponent, $\varLambda_{2}$, $\cdots$, $\varLambda_{n}$, can be estimated similarly, although it is often sufficient to estimate the maximum Lyapunov exponent for clarifying the dynamical state. 
In addition, the Lyapunov exponent estimated here corresponds to a conditional Lyapunov exponent \cite{akashi20}. 

The Lyapunov exponent here is the long-time average of the temporal Lyapunov exponent. 
While the value of the temporal Lyapunov exponent near the initial time ($t=t_{0}$) depends on the choice of the initial perturbation [$\mathbf{m}^{(1)}(t_{0})$],  which is an arbitrary value, the long-time averaged value tends to become a certain value, which is independent of the initial value; see, for example, Ref. \cite{taniguchi19PRB}. 
Because of the finite calculation time, however, the initial value of the temporal Lyapunov exponent might provide some confusion. 
For example, in Fig. \ref{fig:fig7}(b), the label includes a negative value, while we argue that the exponent is zero or positive. 
This is because a negative value in the temporal Lyapunov exponent near the initial time, originated from an arbitrary choice of the initial perturbation, remains. 
As mentioned above, however, it becomes sufficiently small, and the long-time averaged value becomes close to zero. 
We have carefully checked these values and concluded that the Lyapunov exponent in Fig. \ref{fig:fig7}(b) is zero or positive.


\section{Method of evaluating synchronization index}
\label{sec:AppendixC}


\begin{figure*}
\centerline{\includegraphics[width=2.0\columnwidth]{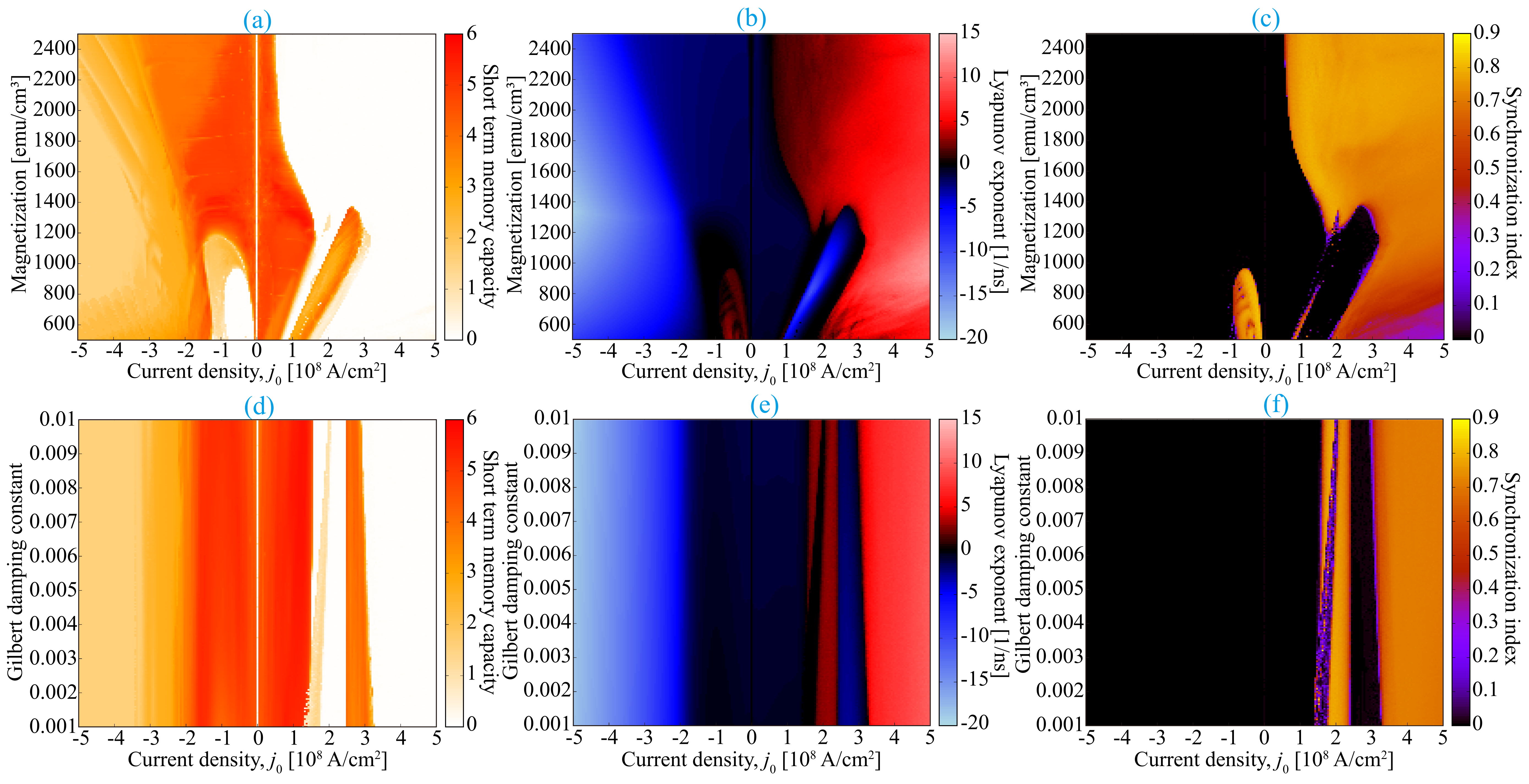}}
\caption{
            (a) Short-term memory capacity, (b) Lyapunov exponent, and (c) synchronization index of an STO with two free layers and one reference layer, where the horizontal axis is the current density and the vertical axis is the saturation magnetization in the F${}_{1}$ layer. 
            (d), (e), and () Same as (a), (b), and (c) a function of the Gilbert damping constant in the F${}_{1}$ layer. 
         \vspace{-3ex}}
\label{fig:fig11}
\end{figure*}

Here, let us summarize the method of evaluating the synchronization index. 
For simplicity, we will suppose an STO with a single free layer as an example, for a while. 

We denote the solutions of the LLG equation with two different initial conditions as $\mathbf{m}^{(1)}$ and $\mathbf{m}^{(2)}$. 
We again emphasize that $\mathbf{m}^{(1)}$ and $\mathbf{m}^{(2)}$ do not have any correlation. 
Then, we evaluate the evolution of their difference. 
Here, the difference is given to the initial state only, while that in Appendix \ref{sec:AppendixB} is given at every time step $t_{n}$. 
Accordingly, $\mathbf{m}^{(a)}$ ($a=1,2$) introduced here has a different meaning from that in Appendix \ref{sec:AppendixB}. 
In the case of STOs with two free layers, we solve the LLG equations for two magnetizations $\mathbf{m}_{i}^{(1)}$ and $\mathbf{m}_{i}^{(2)}$ ($i=1,2$), where there are small differences between the initial states of different samples.

Suppose that a random binary input signal is injected into the STO. 
Therefore, if the STOs for $\mathbf{m}_{i}^{(1)}$ and $\mathbf{m}_{i}^{(2)}$ show noise-induced synchronization \cite{imai22}, the difference will be zero, and the synchronization index will also be zero. 
However, noise-induced synchronization is not the only state that appears in nonlinear oscillators. 
For example, if the STOs originally show chaotic behavior, it is difficult to realize noise-induced synchronization. 
Another possibility is input-driven chaos \cite{akashi20}, where the input signal causes chaos even if the STO does not show chaotic behavior originally. 
In these cases, the synchronization index will remain finite even after a long time passes. 
Therefore, the synchronization index becomes zero only when the distance between the initial states is zero, which indicates the presence of the echo state property, as mentioned in Sec. \ref{sec:Edges of chaos and echo state property}. 

Now let us briefly comment on the definition of the distance $\mathcal{D}$ between the initial states; see also Secs. \ref{sec:Edges of chaos and echo state property} and \ref{sec:Synchronization index}. 
One possible  ``distance'' between two samples of the solution is the relative angle, $\cos^{-1}[\mathbf{m}^{(1)}\cdot\mathbf{m}^{(2)}]$, between two solutions, $\mathbf{m}^{(1)}$ and $\mathbf{m}^{(2)}$. 
This definition relies on the fact that the LLG equation conserves the norm of the solution $\mathbf{m}$, and thus, the magnetization dynamics described by $\mathbf{m}$ with normalization $|\mathbf{m}|=1$ can be regarded as the motion of a point particle on the unit sphere. 
Then, the angle between $\mathbf{m}^{(1)}$ and $\mathbf{m}^{(2)}$ represents their distance measured on the sphere. 
This definition of the distance is used in, for example, Ref. \cite{taniguchi19PRB}. 
For the STOs with two free layers studied in Secs. \ref{sec:Short-term memory capacity, Lyapunov exponent, and synchronization index of STO with two free layers} and \ref{sec:Short-term memory capacity, Lyapunov exponent, and synchronization index of STO with two free layers and one reference layer}, the distance is defined as $\mathcal{D}=\sum_{i=1}^{2} \cos^{-1} \left[ \mathbf{m}_{i}^{(1)}(t)\cdot\mathbf{m}_{i}^{(2)}(t) \right]$, where the suffix $i$ distinguishes the ferromagnetic layers.


A different choice of distance is made in, for example, Refs. \cite{taniguchi20,taniguchi22,taniguchi22JMMM}. 
There, the zenith and azimuth angles, $\theta_{i}^{(a)}$ and $\varphi_{i}^{(a)}$, are $\mathbf{m}_{i}^{(a)}=[m_{ix}^{(a)},m_{iy}^{(a)},m_{iz}^{(a)}]=[\sin\theta_{i}^{(a)}\cos\varphi_{i}^{(a)},\sin\theta_{i}^{(a)}\sin\varphi_{i}^{(a)},\cos\theta_{i}^{(a)}]$, and the distance is defined as $\mathcal{D}=\sqrt{ \sum_{i=1}^{2} \left[ |\theta_{i}^{(1)}-\theta_{i}^{(2)}|^{2} + |\varphi_{i}^{(1)}-\varphi_{i}^{(2)}|^{2} \right]}$. 
In this definition, $\mathcal{D}$ is a distance in a four dimensional phase space consisting of $\theta_{1}$, $\varphi_{1}$, $\theta_{2}$, and $\varphi_{2}$. 

We defined the distance $\mathcal{D}$ differently, as mentioned in Sec. \ref{sec:Synchronization index}. 
The distances in Sec. \ref{sec:Synchronization index}, \ref{sec:Short-term memory capacity, Lyapunov exponent, and synchronization index of STO with two free layers}, and \ref{sec:Short-term memory capacity, Lyapunov exponent, and synchronization index of STO with two free layers and one reference layer} are based on $m_{x}$, $\mathbf{m}_{1}\cdot\mathbf{m}_{2}$, and $m_{1x}$ as $\mathcal{D}=|m_{x}^{(1)}-m_{x}^{(2)}|$, $\mathcal{D}=|\mathbf{m}_{1}^{(1)}\cdot\mathbf{m}_{2}^{(1)}-\mathbf{m}_{1}^{(2)}\cdot\mathbf{m}_{2}^{(2)}|$, and $\mathcal{D}=|m_{1x}^{(1)}-m_{1x}^{(2)}|$, respectively. 
This is because, if these distances tend to be zero, the output signal used for physical reservoir computing becomes independent of the initial state; thus, these distances provide a natural standard with which to study the echo state property for computing.  
It is unnecessary that $\mathbf{m}^{(1)}$ and $\mathbf{m}^{(2)}$ become identical; only the dynamical variable used for the computing should be identical. 
Simultaneously, we note that the difference of the definition of $\mathcal{D}$ in the previous and present paper is important mainly in Sec. \ref{sec:Short-term memory capacity, Lyapunov exponent, and synchronization index of STO with two free layers} only, where even in parameter regions where $\mathbf{m}^{(1)}\neq \mathbf{m}^{(2)}$, $\mathcal{D}=|\mathbf{m}_{1}^{(1)}\cdot\mathbf{m}_{2}^{(1)}-\mathbf{m}_{1}^{(2)}\cdot\mathbf{m}_{2}^{(2)}|$ could be zero, due to the axial symmetry. 
In Secs. \ref{sec:Synchronization index} and \ref{sec:Short-term memory capacity, Lyapunov exponent, and synchronization index of STO with two free layers and one reference layer}, on the other hand, $\mathbf{m}^{(1)}$ and $\mathbf{m}^{(2)}$ [or $\mathbf{m}_{i}^{(1)}$ and $\mathbf{m}_{i}^{(2)}$] become identical in parameter regions where $\mathcal{D}=0$. 
This is because the presence of the reference layer breaks the axial symmetry of the system and reduce the parameter regions where the Lyapunov exponent is zero. 
In summary, the definition of the distance $\mathcal{D}$ should be carefully chosen mainly in Sec. \ref{sec:Short-term memory capacity, Lyapunov exponent, and synchronization index of STO with two free layers} because the system has axial symmetry, due to which, the output signal shows the echo state property even if $\mathbf{m}_{i}^{(1)}\neq \mathbf{m}_{i}^{(2)}$.  


\section{Dependence of short-term memory capacity on parameters in F${}_{1}$ layer}
\label{sec:AppendixD}

In Sec. \ref{sec:Short-term memory capacity, Lyapunov exponent, and synchronization index of STO with two free layers and one reference layer}, the dependence of the short-term memory capacity on the parameters in the F${}_{2}$ layer was studied. 
The parameters, such as the saturation magnetization $M_{2}$ and the damping constant $\alpha_{2}$, can be changed by changing ferromagnetic materials. 
Therefore, we studied the short-term memory capacity by changing them. 
On the other hand, changing the material in the F${}_{1}$ layer might be not preferable because it also changes the magnitude of the output signal generated through the GMR/TMR effect. 
Usually, CoFeB/MgO-based magnetic tunnel junctions are used for STOs, which can emit relatively large power \cite{torrejon17,tsunegi18}. 
However, for comprehensive study, one might be interested in the dependence of the short-term memory capacity on the parameters in the F${}_{1}$ layer. 
Figures \ref{fig:fig11}(a)-\ref{fig:fig11}(c) summarize the short-term memory capacity, the Lyapunov exponent, and the synchronization index of the STO with two free layers and one reference layer, where the vertical axis represents the saturation magnetization $M_{1}$ in F${}_{1}$ layer. 
The other parameters are $M_{2}=1300$ emu/cm${}^{3}$ and $\alpha_{1}=\alpha_{2}=0.01$. 
On the other hand, Figs. \ref{fig:fig11}(d)-\ref{fig:fig11}(f) show the same where the vertical axis represents the damping constant $\alpha_{1}$ in F${}_{1}$ layer while $M_{1}=M_{2}=1300$ emu/cm${}^{3}$ and $\alpha_{2}=0.01$. 
These results indicate that the maximum short-term memory capacity occurs near the edge of the echo state property, which is consistent with the conclusion in the main text.


%



\end{document}